\documentclass[a4paper,12pt]{article}
\usepackage[english]{babel}
\usepackage[utf8]{inputenc}
\usepackage{graphicx} 
\usepackage{fancyhdr} 
\usepackage{amsmath}
\usepackage{physics}
\usepackage{slashed}
\usepackage{dingbat}
\usepackage{cite}
\usepackage{amssymb}
\usepackage{hyperref}

\usepackage{stackengine}
\usepackage{mattens}
\usepackage{eucal}
\usepackage[usenames, dvipsnames]{color}
\numberwithin{equation}{section}

\usepackage{cite}
\begin{document}

\begin{center} 
\vspace*{1.5cm} 
 
{\Large{\textbf {{Mechanisms of Producing Primordial Black Holes  By  Breaking The SU(2,1)/SU(2)$\times$U(1) Symmetry}}}}\\ 
 \vspace*{10mm} 
 
 {\bf  Ioanna~D. Stamou} $^1$ \\ 
  
\vspace{.7cm} 

 $^1${\it National and Kapodistrian University of Athens, Department of Physics, \\
 Section of Nuclear {\rm \&} Particle Physics,  GR--15784 Athens, Greece} \\

\end{center} 
\vspace{2.cm}
 
\begin{abstract}
In this paper we present a class of potentials derived by no-scale supergravity in order to explain the production of primordial black holes (PBHs) in the Universe. By breaking the  SU(2,1)/SU(2)$\times$U(1) symmetry we fix one of the two chiral fields and we derive effective scalar potentials which are capable of generating PBHs. Specifically, we modify well-known superpotentials, which reduce to Starobinsky-like effective scalar potentials. Thus, we derive scalar potentials which, on the one hand, explain the production of PBHs and, on the other hand, they conserve the transformation laws, which yield from the parametrization of the coset SU(2,1)/SU(2)$\times$U(1). Moreover, we generate PBHs by modifying the kinetic term of the Langrangian (or the K\"ahler potential) and we keep the superpotentials unmodified. In all cases we evaluate the fractional abundances of PBHs by comparing Press–Schechter approach and peak theory, while focusing on explaining the dark matter in the Universe. All models are in complete consistence with Planck constraints.

\end{abstract}

\newpage
\section{Introduction}
\label{intro}
\par
\par
One of the greatest problems plaguing Modern Cosmology is that of the origin of the dark matter. A simple explanation is the hypothesis that primordial black holes (PBHs) constitute a significant component of dark matter. This explanation has become a strongly favored object of study after the first detection of gravitational waves (GW) emitted by a binary black hole merge. The detection of this kind reported by LIGO Scientific Collaboration \cite{Abbott:2016blz,Abbott:2016nmj,Abbott:2017vtc,Abbott:2017gyy,Abbott:2017oio} endorses this explanation. As a result of this, many theoretical papers have proposed the connection between dark matter and PBHs \cite{Ballesteros:2017fsr,Ozsoy:2018flq,Biagetti:2018pjj,Franciolini:2018vbk,Gao:2018pvq,Cicoli:2018asa,Dalianis:2018frf,Garcia-Bellido:2017mdw,Ezquiaga:2017fvi,Gong:2017qlj,Hertzberg:2017dkh,Espinosa:2017sgp,Clesse:2015wea,Germani:2017bcs,Inomata:2017okj,Motohashi:2017kbs,Kannike:2017bxn,Inomata:2017vxo,Kawasaki:2016pql,Carr:2016drx,Cheng:2018yyr,Mahbub:2019uhl,Mishra19,Ballesteros:2020qam,Ballesteros:2019hus,Nanopoulos:2020nnh,Braglia:2020eai,Aldabergenov:2020bpt,Anguelova:2020nzl,Braglia:2020taf}.

\par
In the framework of previous theoretical researches, it is suggested that PBHs are generated by reinforcement of density perturbation during inflation, which then collapse and produce PBH at the horizon reentry in the radiation dominated era. Specifically, it is claimed that an inflection point in the effective scalar potential can lead to the formation of a significant peak in the power spectrum which envisages the fractional abundance of dark matter in the Universe. These inflationary potentials should be characterized by a slow-roll period around the time the observable scales exit the horizon, followed by an inflection point region, where the enhancement of power spectrum takes place\cite{Ballesteros:2017fsr,Ozsoy:2018flq,Gao:2018pvq,Cicoli:2018asa,Dalianis:2018frf,Garcia-Bellido:2017mdw,Ezquiaga:2017fvi,Gong:2017qlj,Hertzberg:2017dkh,Mahbub:2019uhl,Mishra19,Ballesteros:2020qam,Ballesteros:2019hus,Nanopoulos:2020nnh}.
\par
Many previous researchers on PBH production from inflation \cite{Ballesteros:2017fsr,Ozsoy:2018flq,Biagetti:2018pjj,Franciolini:2018vbk,Gao:2018pvq,Cicoli:2018asa,Dalianis:2018frf,Garcia-Bellido:2017mdw,Ezquiaga:2017fvi,Gong:2017qlj,Hertzberg:2017dkh,Espinosa:2017sgp,Clesse:2015wea,Germani:2017bcs,Inomata:2017okj,Motohashi:2017kbs,Kannike:2017bxn,Inomata:2017vxo,Kawasaki:2016pql,Carr:2016drx,Cheng:2018yyr,Mahbub:2019uhl,Mishra19,Ballesteros:2020qam,Ballesteros:2019hus,Nanopoulos:2020nnh} alleged that the slow-roll approximation is not regarded as a good evaluation of the power spectrum and for this reason the calculation of the exact equations of fields' perturbations is imperative in order to produce such structures. In recent works there are many models, which improve the prediction of the values of the scalar tilt  $n_s$ and the tensor-to-scalar ratio $r$, such as Refs. \cite{Gao:2018pvq,Ballesteros:2019hus,Nanopoulos:2020nnh,Mishra19} and they are in  in complete consistence with constraints given by Planck measurements of the cosmic microwave background (CMB) anisotropies  \cite{Ade:2015lrj,Akrami:2018odb}.

\par
As it has been mentioned before, the inflationary models needs supersymmetry, as it required an energy scale smaller than the Planck one \cite{Ellis:1982a,Ellis:2013nxa}. In cosmology  it is reasonable to take into consideration inflation in the context of supergravity. The assertion that models inserted in supergravity theories can be notable candidates in order to explain the existence of such peaks has already been studied by many previous works \cite{Gao:2018pvq,Dalianis:2018frf,Mahbub:2019uhl}. In addition to that, it has been claimed that no-scale supergravity can be regarded as a theory worthy of attention because it evades many problematic features, such as the high degree of fine-tuning, which is necessary to obtain a vanishing cosmological constant at the classical level, and the $\eta$ problem\cite{Cremer:1983,Ellislahanas:1984,Elliskounas:1984a,Elliskounas:1984b,Lahanas:1986}. 
\par
 In this paper we develop a method of constructing one field inflationary models by reducing the system from two chiral fields to one single field. We address these models based on the no-scale supergravity model and stay within the constraints of $n_s$ and $r$ by Planck data\cite{Ade:2015lrj,Akrami:2018odb}.  We work in accordance with the symmetries proposed by Refs. \cite{Ellis:20181,Ellis:2013nxa}  based on a non compact coset SU(2,1)/SU(2)$\times$U(1). Moreover, our work is based on a general scheme and we derive significant fractional abundance matching the aspect of no-scale theory and the theory of dark matter origin. First, we modify the superpotential in order, on the one hand, to keep the transformation law presented in Refs.\cite{Ellis:20181,Ellis:2013nxa} intact and, on the other hand, to produce PBH capable to explain the dark matter in the Universe. Second, we present an additional way to produce PBHs by modifying the K\"ahler potential. Therefore, we present a general mechanism in order to produce PBHs.

\par
This paper is organized as follows. In Section 2 we review the general structure of no-scale supergravity. In Section 3 we modify the superpotential based on a general treatment of Starobinsky avatars of no-scale supergravity \cite{Ellis:20181} in order to achieve inflection point in an effective scalar field. In Section 4 we take into consideration the Cecotti superpotential and we modify the K\"ahler potential in order to produce PBH. In  all models we evaluate the power spectrum both via the slow-roll approximation and by solving the exact equations of field's perturbation. Furthermore, in order to evaluate the fractional abundances, we make a comparison between the Press–Schechter approach (PS)\cite{Press74} and peak theory (PT) \cite{Bardeen:1985tr}.  In Section 5 we discuss the fine-tuning issue. Finally, in Section 6 we present our conclusions and remarks. 
\section{Supergravity framework}
  \label{ballesterpbh}
 In this section we provide some general features of no-scale supergravity which will be requisite for our analysis. First, we recall some properties of the $\mathcal{N}=1$ supergravity Langrangian coupled to matter: 

\begin{equation}
\label{01}
\mathcal{L}=  K^{\bar{j}}_{i} \partial_{\mu} \Phi^i \partial ^{\mu} \bar{\Phi} _{\bar{j}}-V(\Phi)
\end{equation}
\noindent
with $i,j$ running over all fields. We denote with $\Phi^i$  the chiral fields and $ \bar{\Phi} ^{\bar{j}}$ the conjugate fields. The K\"ahler metric in the kinetic term is given by:
\begin{equation}
K^{\bar{j}}_{i}= \frac{\partial^2K}{\partial \Phi^i \partial \bar{\Phi}_{\bar{j}}}
\label{02}
\end{equation}
\noindent
and $K$ represents the K\"ahler potential. The scalar potential $V$ can be expressed in terms of the K\"ahler potential, $K$ and the superpotential $W$:
\begin{equation}
V= e^{K/{M_P}^2}\left[\left( {K^{-1}}\right)^i_{\bar{j}} \left(W^{\bar{j}}+\frac{WK^{\bar{j}}}{{M_P}^2}\right) \left(\bar{W}_i+\frac{\bar{W}K_{i}}{{M_P}^2}\right)- \frac{3|W|^2}{{M_P}^2} \right]+\frac{\bar{g}}{2}\left(K^iT^{\alpha}	\Phi_i\right)^2
\label{3}
\end{equation}
\noindent
where $( {K^{-1}})^i_{\bar{j}}$ is the inverse of K\"ahler metric and $K^i=\partial K/ \partial \Phi_i$.  The term proportional to $e^{K/{M_P}^2}$ is the $F$-term of the potential whereas the term proportional to $\bar{g}$ is the $D$-term. We set $\bar{g}=0$, since the $D$-term is absent because chiral fields are gauge singlet. One can rewrite the above equation in a more compact form using the covariant derivatives:
\begin{equation}
\mathcal{D}_i W=\partial_i W+ \frac{1}{{M_P}^2}K_i W.
\label{3con}
\end{equation}
We work in Planck units ($M_P=1$) throughout this paper, where $M_P$ is the reduced mass Planck.

\par
The simplest example of no-scale models can be written as \cite{Cremer:1983,Ellislahanas:1984,Elliskounas:1984a,Elliskounas:1984b,Lahanas:1986}:
\begin{equation}
\label{3i}
K=-3 \ln( \Phi+ \bar{\Phi} )
\end{equation} 
\noindent
depending on a single chiral field $\Phi$. If we consider the scalar potential (\ref{3}) the term $-3|W|^2$ vanishes due to the identity \cite{Cremer:1983}: 
\begin{equation}
\label{nscont}
K^{\Phi \bar{\Phi}}K_\Phi K_{\bar{\Phi}}=3. 
\end{equation}
\noindent
The K\"ahler metric of (\ref{3i}) is $ 3/(\Phi+ \bar{\Phi})^2$ and corresponds to the Ricci tensor $\mathcal{R}^i_{\bar{j}}=\frac{2}{3}K^i_{\bar{j}} $. Hence the K\"ahler curvature reduces to the Ricci scalar $\mathcal{R}=2/3$. This K\"ahler manifold can be identified with the coset SU(1,1)/U(1), which is invariant under the SU(1,1) group of isometric transformation \cite{Elliskounas:1984a} :
\begin{equation}
\Phi \rightarrow \frac{\alpha \Phi+i \beta}{i \gamma \Phi +\delta}, \quad \alpha,\beta,\gamma,\delta \in \rm \mathbb{R}\quad \text{with}  \quad \alpha\delta+\beta\gamma =1.
\end{equation}
\par
A more reviling form of SU(1,1) symmetry is determined, if we define the complex field $y$ as \cite{Elliskounas:1984b,Ellis:2013nxa}:
\begin{equation}
\Phi =\frac{1}{2} \left (   \frac{1-y/ \sqrt{3}}{1+ y/ \sqrt{3}}  \right).
\end{equation}
\noindent
Then the K\"ahler potential takes the form:
\begin{equation}
K=-3\ln\left (   1-\frac{|y|^2}{3}  \right ) .
\end{equation}
\noindent
The SU(1,1)/U(1) coset space can be parametrized by the $2 \times 2$ complex matrix ${U^\dag }JU$ with:
\begin{equation}
U=\begin{bmatrix}
A & B \\
\bar{B} & \bar{A} 
\end{bmatrix} 
\quad \text{where} \quad
J=\begin{bmatrix}
1 & 0 \\
0 & -1 
\end{bmatrix} \text{and}\quad  A,B  \in \rm \mathbb{C} .
\end{equation}
\noindent
Thus the Langrangian is invariant under the transformation law:
\begin{equation}
y \rightarrow  \frac{A y +B}{\bar{B}y +\bar{A}} \quad \text{with} \quad |A|^2-|B|^2=1.
\end{equation}

The minimal no-scale SU(1,1)/U(1) model cannot lead to the Starobinsky effective scalar potential,  because there are not such superpotentials to support it \cite{Ellis:2013nxa}. But, we can find such superpotentials, if we consider an additional modulo chiral field \cite{Ellis:2013xoa,Ellis:2013nxa,Ellis:20181}. The  SU(1,1)/U(1) coset is extended to the SU(2,1)/SU(2)$\times$ U(1) one and  the K\"ahler potential, which can be parametrized by this coset, is given from the equation \cite{Elliskounas:1984b}:

\begin{equation}
K=-3 \ln( T+ \bar{T} - \frac{\varphi \bar{\varphi}}{3})
\label{1}
\end{equation} 
\noindent
where $T$, $\varphi$ are two chiral superfields.

In the extension of $\mathcal{N} =2$ no-scale supergravity there is the choice of which field will take the role of inflaton and which will be regarded as a modulo field. A possible superpotential, notable to be considered, was presented in Ref.~\cite{Ellis:2013nxa} and is the simplest Wess-Zumino model given by the equation:
 \begin{equation}
W= \frac{\hat \mu}{2} \varphi^2  - \frac{\lambda}{3} \varphi^3
 \label{1a}
 \end{equation}
where $\mu$ is the mass term and $\lambda$ is a trilinear coupling. This superpotential is characterized by one single chiral superfield. We can divulge the effective scalar Starobinsky potential by considering the field $\varphi$ as the inflaton field and the field $T$ as the modulo field. In the direction  $T= \bar{T}= 1/2$ and $Im\varphi=0$ if we set $\lambda / \mu = 1/3$ and $\hat{\mu}=\mu \sqrt{1/3}$, we can reduce the scalar potential to the Starobinsky one. In order to achieve this, one should fix the non canonical kinetic term in the Langrangian by a redefinition of the field:
\begin{equation}
\varphi= \sqrt{3} \tanh (\frac{\chi }{\sqrt{6}}).
\end{equation}
Therefore by using the Eq.~(\ref{3}) one can find the effective scalar potential:
\begin{equation}
V=\mu^2 e^{-\sqrt{2/3}\chi}\sinh^2\left( \frac{\chi}{\sqrt{6}}\right).
\end{equation}

Another possible scenario in order to retrieve the Starobinsky effective scalar potential by no-scale theory is shown in Ref. \cite{Cecotti:1987a}. The superpotential proposed is:

\begin{equation}
W= \sqrt{3}m \varphi \left(T- \frac{1}{2}\right)
\label{2}
\end{equation} 
where the $T$ plays the role of inflaton and $\varphi$ is the modulo field. The Starobinsky potential is concealed in the direction $\varphi=  \bar{\varphi}=\Im T=0$ and the proper redefinition of the field in order to have canonical kinetic terms is:
\begin{equation}
T=\frac{1}{2}e^{\sqrt{\frac{2}{3}}\chi}.
\end{equation}
\noindent
As before, the effective scalar potential is found by using Eq.~(\ref{3}):
 \begin{equation}
 V(\chi)=\frac{3m^2}{4}\left(1- e^{-\sqrt{\frac{2}{3}}\chi}\right)^2.
 \label{staro}
 \end{equation}

Up until now we have shown the construction of Starobinsky-like potentials by a K\"ahler potential and a superpotential. However, a complete presentation of these desirable effective scalar potentials should always include the confirmation of the stabilization in the direction of interest. Therefore, one should compute the mass matrix of the full Langrangian in order to verify that there are no tachyonic instabilities. Therefore, as we modify the superpotential and K\"ahler potential, we always have to check the stabilization by using the methodology described in Refs. \cite{Ellis:2019bmm,Kallosh:2010ug,Ellis:2013nxa}.

\section{PBH production in the SU(2,1)/SU(2)$\times$U(1) symmetry framework}
  \label{attr}
We study potentials which can generate PBHs that might account for a significant fraction of the dark matter in the Universe. These potentials should be characterized by an inflection point where the inflaton slows down, producing a quick rise in curvature perturbation. This feature is explained as a region of ultra slow-roll, where meaningful spikes are produced. We can achieve such inflection points by modifying the superpotentials given in Eqs~(\ref{1a}) and (\ref{2}).  First, as we will see, we modify the superpotential in $(y_1,y_2)$ basis and secondly in $(T,\varphi)$ basis \cite{Ellis:20181,Ellis:2013nxa}. The purpose of this section is finding a class of proper potentials which have such features.

We consider the simplest nonminimal no-scale supergravity model based on the SU(2,1)/ SU(2)$\times$U(1) coset. There are two equivalent forms of K\"ahler potential, which can parametrize this coset manifold \cite{Ellis:20181}:
\begin{equation}
K=-3\ln(1-\frac{|y_1|^2}{3}-\frac{|y_2|^2}{3}) \quad or \quad K=-3 \ln(T +\bar{T}- \frac{|\varphi|^2}{3}).
\label{eq:2k}
\end{equation}
As Ref. \cite{Ellis:20181} mentions one can transform between these two equal forms using the non compact SU(2,1)/SU(2)$\times$U(1) symmetry.

\begin{figure}[h!]
\centering
\includegraphics[width=155mm]{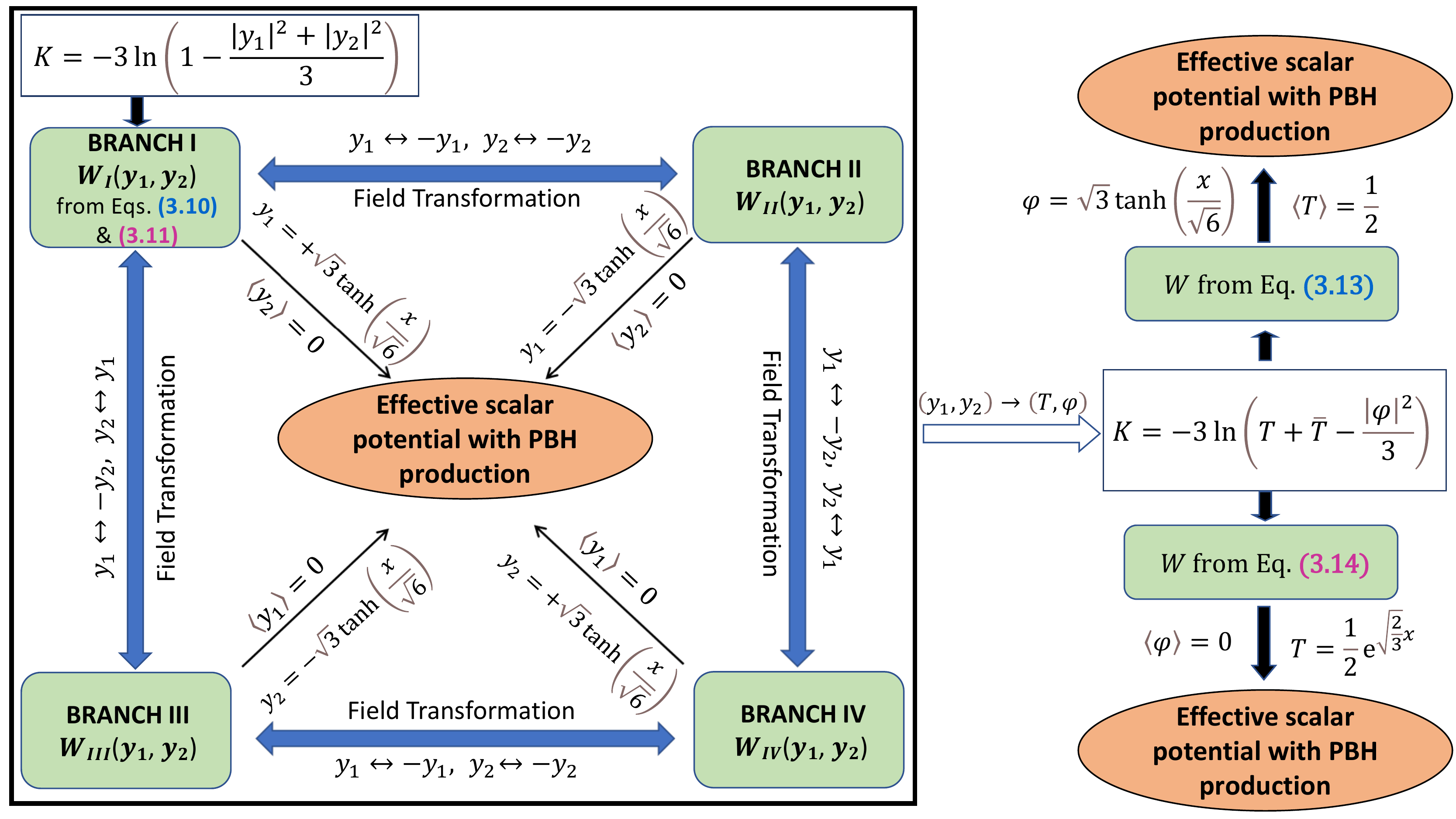} 
\caption{Diagram, adapted from Ref~ \cite{Ellis:20181}. (For the superpotential to go from branch I to the other ones we start from Eqs.~(\ref{k2(8)}) and (\ref{k2(13)}) and apply the transformations as shown).}
\label{fbran}
\end{figure}

In this analysis, we consider K\"ahler potentials given by Eq.~(\ref{eq:2k}) and we use modified superpotentials, as they are presented in Figure \ref{fbran}, in order to find suitable results in PBH scenario. This figure, based on the analysis of Refs.~\cite{Ellis:20181}, shows that starting from a general two field $W(T,\varphi)$ or $W(y_1,y_2)$, it is possible to break the SU(2,1)/SU(2)$\times$U(1) symmetry by fixing one of the complex scalar fields. One should calculate the redefinition of the fields depicted in Figure~\ref{fbran} in order to fix the non canonical kinetic term and proceed to the evaluation of the Starobinsky-like effective scalar potential.  In addition, in Figure~\ref{fbran} we present that, in the basis $(y_1,y_2)$, starting by one superpotential each time, we can conserve some symmetries.

To begin with, we adopt a more symmetric representation of SU(2,1) /SU(2)$\times$U(1) coset by choosing the basis $(y_1,y_2)$ \cite{Elliskounas:1984b,Ellis:2013nxa,Ellis:20181}. Hence the  K\"ahler potential takes the form:
\begin{equation}
K=-3 \ln\Big(1-\frac{|y_1|^2}{3}-\frac{|y_2|^2}{3}\Big)
\label{k2(1)}
\end{equation}
 \noindent
where the complex fields $(y_1,y_2)$ are related to $(T,\varphi)$ appearing in (\ref{1}) by the following expressions:
 \begin{equation}
 y_1=\Big(\frac{2\varphi}{1+2T}\Big) ,\quad y_2=\sqrt{3}\Big(\frac{1-2T}{1+2T}\Big)
 \label{k2(2)}
 \end{equation}
 \noindent
 and the inverse relations by:
  \begin{equation}
 T=\frac{1}{2}\Big(\frac{1 -y_2 / \sqrt{3}}{1 +y_2 / \sqrt{3}}\Big) ,\quad \varphi=\Big(\frac{y_1}{1 +y_2 / \sqrt{3}}\Big).
 \label{k2(3)}
 \end{equation}
 \noindent
 The superpotential transforms as \cite{Ellis:2013nxa}:
 \begin{equation}
 W(T,\varphi) \rightarrow \bar{W}(y_1,y_2)= (1+y_2 /\sqrt{3})^3W.
  \label{k2(4)}
 \end{equation}
  \noindent
By this redefinition the Wess-Zumino superpotential given in Eq.~(\ref{1a}) becomes \cite{Ellis:2013xoa}:
\begin{equation}
W=\frac{\hat\mu}{2} \Big(y_1^2 +\frac{y_1^2 y_2}{\sqrt{3}}\Big) -\lambda \frac{y_1^3}{3}.
\label{k2(5)}
\end{equation} 
and the Cecotti superpotential in (\ref{2}) is modified \cite{Cecotti:1987a,Kallosh:2013xya}:
\begin{equation}
W=m\Big( -y_1y_2 +\frac{y_2y_1^2}{\sqrt{3}}\Big ).
\label{k2(6)}
\end{equation}
The SU(2,1)/SU(2)$\times$U(1) coset space is parametrized by the matrix $U$:
\begin{gather}
U
 =
  \begin{bmatrix}
   \alpha &
  \beta & 0 \\
   -\beta^* & \alpha^* & 0 \\
   0& 0& 1 
   \end{bmatrix}
\end{gather}
\noindent
where $\alpha,\beta \in\mathbb{C} $ and $|\alpha|^2+|\beta|^2=1$. By this matrix and the analysis shown in Ref.~\cite{Ellis:20181} we can obtain the following transformation laws for the fields:
\begin{equation}
y_1 \rightarrow \alpha y_1 + \beta y_2, \quad y_2 \rightarrow - \beta^* y_1 +\alpha ^* y_2.
\label{k2(7)}
\end{equation}
\noindent
Applying this transformation the K\"ahler potential remains invariant, whereas the superpotential $W(y_1,y_2)$ is modified nontrivially.  In the case of superpotentials Eqs.~(\ref{k2(5)}) and (\ref{k2(6)}) these transformations  are invariant for proper values of $\alpha$ and $\beta$, as shown by Ref.~\cite{Ellis:20181}. 
In Figure \ref{fbran} we show that the conditions (\ref{k2(7)}) are satisfied and under which transformation of the field we can derive potentials with canonical kinetic terms. As it was mentioned, the superpotential given from (\ref{k2(5)}) and (\ref{k2(6)}) can satisfy these transformation laws and lead to Starobinsky-like potentials. We notice that we begin by the branch I of Figure~\ref{fbran} and we derive the same effective scalar potential.

We show that, by a proper modification of these superpotentials (see Eqs.~(\ref{k2(5)}) and (\ref{k2(6)})), not only can we  keep the transformation laws illustrated in Figure~\ref{fbran}, but we can also produce PBH.
Therefore, we modify the superpotential (\ref{k2(5)}) as:
\begin{equation}
W=\Big(\frac{\hat \mu}{2} \Big(y_1^2 +\frac{y_1^2 y_2}{\sqrt{3}}\Big) -\lambda \frac{y_1^3}{3}\Big ) (1+g(y_1))
\label{k2(8)}
\end{equation}
\noindent
where $g(y_1)=e^{-b_1y_1^2}(c_1 y_1^2+c_2y_1^4)$. We choose the $y_1$ direction because  Starobinsky-like effective scalar potentials occur at this direction. After the change of basis $(T,\varphi) \rightarrow(y_1,y_2)$ in the  Wess-Zumino superpotential, given by  Eqs.~(\ref{1a}) and (\ref{k2(5)}), the Starobinsky scalar potential arises in the $y_1$ direction. In the following the modification takes place in the field where the Starobinsky effective scalar potential arises in the unmodified case.
%

\begin{figure}[h!]
\centering
\includegraphics[width=1.1\textwidth]{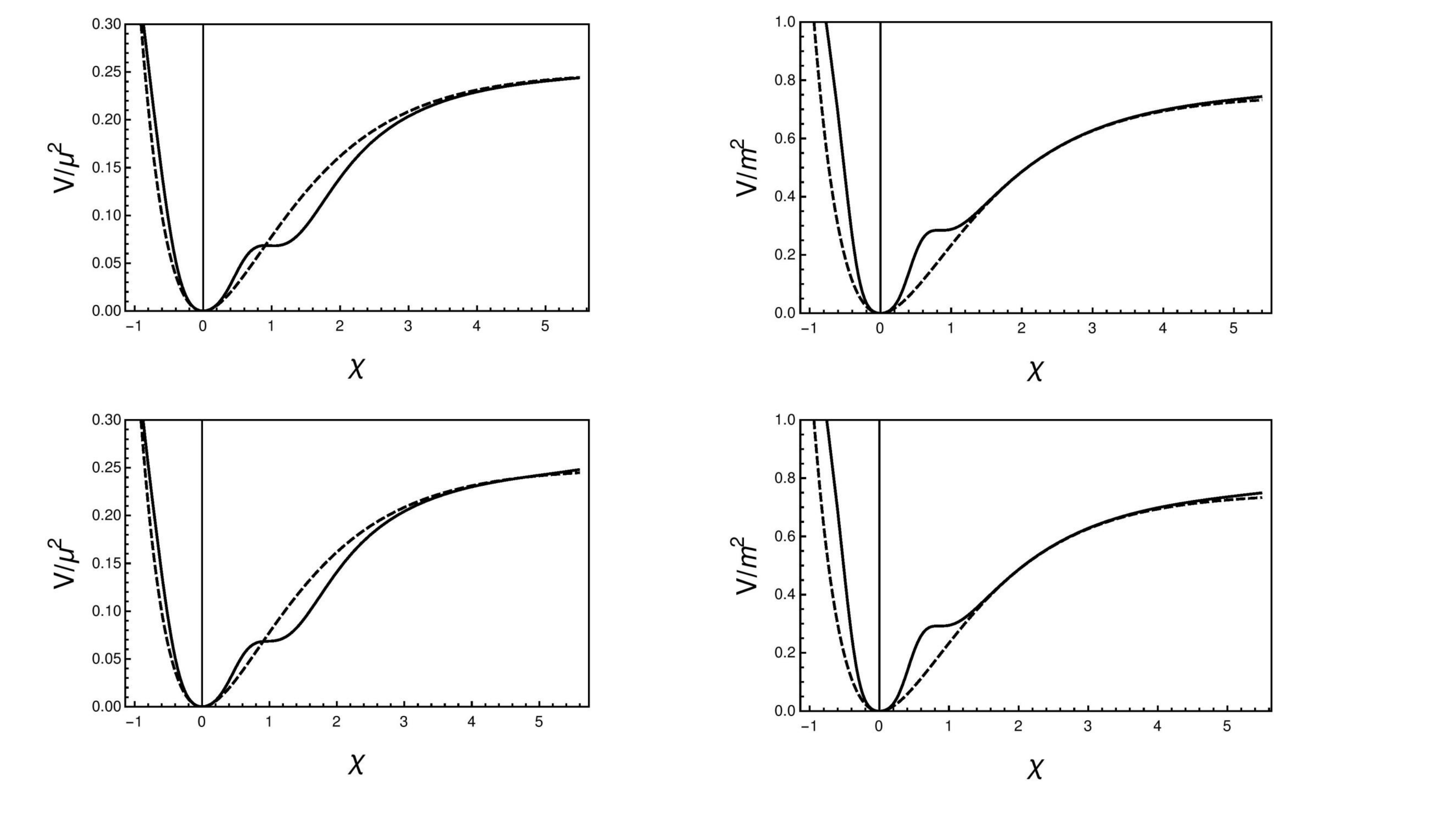}
\caption{Effective scalar potential derived from the superpotentials Eq.~(\ref{k2(8)}) (up left), Eq.~(\ref{k2(13)})(up right),  Eq.~(\ref{wcase3}) (down left) and Eq.~(\ref{wcase4}) (down right). Dashed lines corresponds to Starobinsky potential. In left plots we use the first set of parameters in Table~\ref{tkwz1} and in the right plots the first in Table~\ref{tkcc1}.  }
\label{f3}
\end{figure}

One can see that the choice of $g(y_1)$ does not affect the transformation laws and hence we can derive the same effective scalar potential capable of reproducing PBH throughout the transformation laws depicted in Figure~\ref{fbran}. In Figure~\ref{f3} we show  the effective scalar potential (top left panel), which is evaluated by Eq.~(\ref{3}) using the first set of parameters presented in Table~\ref{tkwz1} and fixing the canonical kinetic term as shown in Figure~\ref{fbran}.

Furthermore, we modify the Cecotti superpotential given from Eq.~(\ref{k2(6)}) as:
\begin{equation}
W=m\Big(-y_1y_2 +\frac{y_2y_1^2}{l\sqrt{3}}\Big)(1+c_3e^{-b_2y_1^2}y_1^2).
\label{k2(13)}
\end{equation}
\noindent
We notice that, by the transformation laws between $y_1$ and $y_2$ depicted in Figure~\ref{fbran}, we can always derive the same effective scalar potential suitable to generate PBH. We use the first set of parameters in Table~\ref{tkcc1} and we plot the potential in Figure~\ref{f3} (top right panel). We begin as before from the branch I in Figure~\ref{fbran} and we remark that for all of our cases we take into consideration the field transformation in order to obtain the canonical kinetic terms. All these transformations are shown in Figure~\ref{fbran}.

We can derive similar results by changing the basis $(y_1,y_2)$ to $(T,\varphi)$. Using the Eq.~(\ref{k2(2)}) and the reverse  Eq.~(\ref{k2(4)}), we can have the K\"ahler potential and superpotential in the basis $(T,\varphi)$. Specifically the K\"ahler potential takes the standard form:
\begin{equation}
K=-3 \ln(T +\bar{T}- \frac{|\varphi|^2}{3})
 \label{kahler}
\end{equation}
In addition to that the superpotential coming from the modification of the Wess-Zumino one Eq.~(\ref{k2(8)}) takes the following form, in the $(T, \varphi)$ basis by using  Eq.~(\ref{k2(2)}) and the reverse  form of Eq.~(\ref{k2(4)}):
\begin{equation}
W=\left( \frac{\hat{\mu}}{2}\varphi^2 -\frac{\lambda}{3} \varphi^3\right)\left(  1+ e^{-b_1 \left( \frac{2 \varphi}{1+ 2 T} \right)^2} \left[ c_1 \left( \frac{2 \varphi}{1+ 2 T} \right)^2+ c_2 \left( \frac{2 \varphi}{1+ 2 T} \right)^4\right] \right)
\label{wcase3}
\end{equation}
We redefine the field $\varphi$ in order to fix the kinetic term as shown in Figure~\ref{fbran}. If we set $c_1=c_2=0$, we have the Wess-Zumino superpotential and we derive the Starobinsky potential, as presented in Ref.~\cite{Ellis:2013xoa}. In Figure \ref{f3}, in the down left plot, we show the effective scalar  potential using the first set of parameters from Table \ref{tkwz1}. We can notice that the inflection point is conserved after the change of the basis.

Moreover, the superpotential, derived by the modification of Ceccoti Eq. (\ref{k2(13)}) reads in the $(T, \varphi)$ basis  as:
\begin{equation}
W= \frac{\sqrt{3}}{2} m \varphi \Big( \frac{1}{2} -T\Big) \left( -1-2T -\frac{1-2T}{l} \right) \left[ 1+ 3c_3 e^{-3b_2 \left( \frac{1-2T}{1+2T} \right)^2}\left( \frac{1-2T}{1+2T} \right)^2\right]
\label{wcase4}
\end{equation}
From this superpotential we evaluate the effective scalar potential by the redefinition of the field as shown in   Figure~\ref{fbran}. We plot the result in down right panel of  Figure \ref{f3} by using the first set of parameter in Table \ref{tkcc1}. As one can notice we derive the same potential, as in the case of Eq. (\ref{k2(13)}). We derive the Starobinsky effective scalar potential when $c_3=0$ and $l=1$. 
\begin{table}[h!]
\centering
 \begin{tabular}{|c|c|c|c|c|}
\hline
 & $c_1$ &$c_2$& $b_1$  &$\mu(10^{-5})$\\
\hline
1 & $2.0$ & $-0.80$&$4.211709$  & $3.29$\\
\hline
2 & $2.0$ & $-0.76$&$4.0906628$  & $3.3$\\
\hline
3 & $2.4$ & $-0.95$&$4.545369$  & $3.53$\\
\hline
4 & $2.3$ & $-0.90$&$4.4365369$  & $3.55$\\
\hline
\end{tabular}
 \caption{ Parameters for the superpotentials (\ref{k2(8)}) and (\ref{wcase3}) with $\lambda / \mu=1/3$ and $\hat \mu =\mu / 
 \sqrt{3}$.}
 \label{tkwz1}
\end{table}

\begin{table}[h!]
\centering
 \begin{tabular}{|c|c|c|c|c|}
\hline
 & $c_3$ & $b_2$  &$\l$ & $m(10^{-5})$\\
\hline
1 & $14$ & $8.99572$&$1.0002$& $2.25$ \\
\hline
2 & $13.5$ & $8.7341953$&$1.00025$  & $2.484$\\
\hline
3 & $13$ & $8.469$&$1.00018$ & $2.15$\\
\hline
4 & $15$ & $9.50485$&$1.00015$ & $2.08$\\
\hline
\end{tabular}
 \caption{ Parameters for the superpotentials (\ref{k2(13)}) and (\ref{wcase4}).}
 \label{tkcc1}
\end{table}

Before we continue with the analysis of evolution of the field, it is imperative to discuss the stabilization in the inflationary direction. We have already considered four cases:
\begin{itemize}
\item Case 1: The K\"ahler potential and superpotential  given from Eqs.~(\ref{k2(5)}) and (\ref{k2(8)}):
\begin{equation}
\begin{split}
 K=-3 \ln\left(1-\frac{|y_1|^2}{3}-\frac{|y_2|^2}{3}\right),\\
W=\left(\frac{\hat\mu}{2} \left(y_1^2 +\frac{y_2^2 y_1}{\sqrt{3}}\right) -\lambda \frac{y_1^3}{3}\right ) \left(1+e^{-b_1y_1^2}(c_1 {y_1}^2+c_2{y_1}^4)\right)
\end{split}
 \label{case1a}
\end{equation}
\noindent
where we have fixed the modulo field and the inflaton through the circle illustrated in Figure~\ref{fbran} starting by branch I. We notice that the same effective scalar potential will arise.
\item  Case 2: The K\"ahler potential and superpotential  given from Eqs. (\ref{k2(5)}) and (\ref{k2(13)}):
\begin{equation}
\begin{split}
 K=-3 \ln\left(1-\frac{|y_1|^2}{3}-\frac{|y_2|^2}{3}\right),\\
 W=m\left(-y_1y_2 +\frac{y_2y_1^2}{l\sqrt{3}}\right)\left(1+c_3e^{-b_2{y_1}^2}{y_1}^2\right)
 \end{split}
 \label{case2}
\end{equation}
\noindent
where we have followed the Figure~\ref{fbran} as before. 

\item Case 3: The K\"ahler potential and superpotential  given from Eqs.~(\ref{1}) and (\ref{wcase3}):
\begin{equation}
\begin{split}
 K=-3 \ln(T +\bar{T}- \frac{|\varphi|^2}{3}),\\  
 W=\left( \frac{\hat{\mu}}{2}\varphi^2 -\frac{\lambda}{3} \varphi^3\right)\left(  1+ e^{-b_1 \left( \frac{2 \varphi}{1+ 2 T} \right)^2} \left[ c_1 \left( \frac{2 \varphi}{1+ 2 T} \right)^2+ c_2 \left( \frac{2 \varphi}{1+ 2 T} \right)^4\right] \right)
 \end{split}
 \label{case3}
\end{equation}
\noindent
where we assume $T=\frac{1}{2}$ and $\Im\varphi=0$.
\item Case 4: The K\"ahler potential and superpotential given from Eqs.~(\ref{1}) and (\ref{wcase4}):
\begin{equation}
\begin{split}
K=-3 \ln(T +\bar{T}- \frac{|\varphi|^2}{3}),\\
W= \frac{\sqrt{3}}{2} m \varphi \Big( \frac{1}{2} -T\Big) \left( -1-2T -\frac{1-2T}{l} \right) \left[ 1+ 3c_3 e^{-3b_2 \left( \frac{1-2T}{1+2T} \right)^2}\left( \frac{1-2T}{1+2T} \right)^2\right]
\end{split}
  \label{case4}
\end{equation}
\noindent
where we assume $\varphi=\Im T=0$.

\end{itemize}
\noindent
We need to show that we do not have tachyonic instabilities along the inflationary direction. We follow the methodology presented in \cite{Ellis:2019bmm}. The general form of mass matrix is given by \cite{Ellis:2019bmm,WessBagger}:
\begin{equation}
m^2_{s}=\begin{pmatrix} 
(K^{-1})^i_k \mathcal{D}^k \partial_j V & (K^{-1})^i_k \mathcal{D}^k \partial^j V \\
 (K^{-1})^k_i \mathcal{D}_k \partial_j V  & (K^{-1})^k_i \mathcal{D}_k \partial^j V
\end{pmatrix} \,, 
\label{massmatrix}
\end{equation} 
\noindent
where the  inverse  K\"ahler  metric is given in Eq.~(\ref{02}) and the covariant derivative by Eq.~(\ref{3con}). We need to diagonalize the matrix (\ref{massmatrix}), in order to determine that we do not have instabilities in the direction of interest. We take into consideration both real and imaginary scalar fields. We confirm that all of our cases are stabilized for the sets of parameters presented above.  Therefore, we achieve to avoid the problem of tachyonic solution   and we study the simple case of single field.

To sum up, we draw some aspects about the cases in Eqs.~(\ref{case1a}), (\ref{case2}), (\ref{case3}) and (\ref{case4}). First, all these cases   satisfy the condition (\ref{nscont}).
We also remark that the  SU(2,1)/SU(2)$\times$U(1) symmetry is defined by a K\"ahler  potential with two equivalent forms, as shown in Eq.(\ref{eq:2k}), for a general superpotential. This K\"ahler  remains invariant under the transformation laws of  Eq.(\ref{k2(7)}), which define the coset SU(2,1)/SU(2)$\times$U(1) and the superpotential is modified nontrivially.  We use two superpotentials adopted from the literature \cite{Ellis:2013nxa,Ellis:20181}, which are given in Eqs. (\ref{k2(5)}) and  (\ref{k2(6)}), and   under these transformations (for the proper value of $\alpha$ and $\beta$ of  Eq.(\ref{k2(7)})) give the same effective Starobinsky-like scalar potential  \cite{Ellis:20181}. Then we modify these properly, in order, not only to keep that symmetry intact, meaning that the same effective scalar potential throughout the branches of Figure~\ref{fbran} arises, but also to explain the production of PBH in the Universe.
 Moreover, the cases  (\ref{case1a}) and (\ref{case2}) keep the transformation laws for the fields depicted in Figure~\ref{fbran}.  In Figure~ \ref{f3} we notice that we achieve to derive an inflection point in all shown graphs of effective scalar potentials. Hence, we accomplish the necessary conditions in order to produce PBH, as it is proposed in Refs.~\cite{Ballesteros:2017fsr,Espinosa:2017sgp,Nanopoulos:2020nnh,Dalianis:2018frf}.

We need to remark that the potentials are sufficiently flat in the CMB limit. However, it is not flat for large values of the field. As it is  mentioned in Ref. \cite{Mishra19}, a base potential with higher value of scalar tilt $n_s$ becomes strongly favored when it is modified with an inflection point. We will present a model with a flat potential at large field values later.

\subsection{Background equation}

In the previous section we presented effective scalar potentials with inflection points. In this section we analyze the evolution of the field and we show that these inflection points leads to the violation of slow-roll approximation. Specifically, the inflection point makes the inflaton to enter a nonattractor phase wasting a few e-folds, which are capable to give an important enhancement in power spectrum. This enhancement is supported by solutions based on slow-roll approximations. However, as we will show, this approximation is violated and hence its results are not reliable.

  In general, the dynamics of a single scalar field in presence of an effective scalar potential $V(\chi)$ is derived by minimizing the inflaton action with respect to $\chi$ and by producing the equation of motion or the background equation. It is more convenient to express the evolution of the field in terms of number of e-folds $N$. Hence the background equation is as follows:
\begin{equation}
\label{21}
\frac{d^2 \chi}{d N^2} +3 \frac{d \chi}{dN}- \frac{1}{2} \left(\frac{d \chi}{d N}\right)^3 +\left(3- \frac{1}{2} \left(\frac{d \chi }{d 	N}\right)^2\right) \frac{d\ln V(\chi)}{d \chi}=0.
\end{equation}  
In this notation the Friedmann equation is as follows:
\begin{equation}
 \label{23}
H^2=\frac{V}{(3- \varepsilon_H)}.
\end{equation}
\noindent
Moreover, we define the slow-roll parameters $\varepsilon_H$ and $\eta_H$:
\begin{equation}
 \label{22}
\varepsilon_H=\frac{1}{2} \left( \frac{d \chi}{dN} \right)^2, \quad \eta_H= \varepsilon_H- \frac{1}{2} \frac{d \ln   \varepsilon_H}{dN}
\end{equation}
and we notice that inflation demands $\varepsilon_H<1$. 
The necessary condition of validity of slow-roll is given by:
\begin{equation}
\varepsilon_H\ll 1, \quad \eta_H \ll 1.
\end{equation} 
\noindent
The slow-roll parameters, $\varepsilon_H$ and $\eta_H$, can also be described as parameters of the inflationary potential, named $\varepsilon_V$ and $\eta_V$ respectively:
\begin{equation}
\varepsilon_V= \frac{1}{2}\left(\frac{V'(\chi)}{V(\chi)} \right)^2 , \quad \eta_V=\frac{V''(\chi)}{V(\chi)}.
\end{equation}
where with primes we denote the derivatives in respect of the field $\chi$.

The Eq.(\ref{21}) can be solved numerically until inflation ends, meaning that the condition $\varepsilon_H=1$ is satisfied.  Regarding that the initial time coincides with the Hubble crossing condition $k= a H$, where $a$ is the scale factor, we derive proper values for $n_s$ and $r$ in the slow-roll approximation by the observational constraints of \cite{Ade:2015lrj,Akrami:2018odb,Akrami2018}.

We work in compatibility with current observational constraints from Planck by calculating $n_s$ and $r$ at leading order in the slow-roll expansion by:
\noindent
\begin{equation}
\label{21a}
n_s \simeq 1+2 \eta_V -6\varepsilon_V, \quad r \simeq 16 \varepsilon_V.
\end{equation} 
We just need an initial condition for the field $\chi$, as the derivative of $\chi$ is founded by supposing initially that we are in the slow-roll attractor: $\frac{d \chi}{dN} \approx -\left|\frac{1}{V}{\frac{dV}{ d\chi}}\right|$. For this reason we set as initial condition those which represent an almost flat region of the potential.
\noindent

We solve Eq.~(\ref{21}) for all cases we presented in the previous section derived from the expressions (\ref{case1a}), (\ref{case2}), (\ref{case3}) and (\ref{case4}). For the following analysis, we use the first set of parameters given in Tables~ \ref{tkwz1} and \ref{tkcc1}   for cases (\ref{case1a}), (\ref{case2}) and second set of parameters given in Tables~\ref{tkwz1} and \ref{tkcc1} for cases (\ref{case3}), (\ref{case4}) respectively. 
 
The inflaton experiences a slow-roll evolution at CMB scales. Because of this stage of evolution we can calculate the observable parameters $n_s$ and $r$, shown in Table~\ref{tbrans} and remark the compatibility with the Planck 2018 constraints. After this stage, the inflaton approaches the inflection point and it decelerates depicted by a plateau. When the field approaches the global minimum, the inflation ends and the stage of reheating takes place illustrated by the small oscillation in evolution of the field.  In the Table~\ref{tbrans} we also show the initial conditions of the field for each case. The $\chi_0$ is the value corresponding to $k=0.05Mpc^{-1}$.

\begin{table}[h!]
\centering
 \begin{tabular}{|c|c|c|c|}
\hline
 & $\chi_0$ & $n_s$& $r$ \\
\hline
1 & $4.98$ &$0.9602$ &$0.0085$  \\
\hline
 2 & $4.89$ &$0.9617$ &$0.0125$\\
\hline
3 & $5.01$ &$0.9622$ &$0.0083$  \\
\hline
 4 &$4.84$ &$0.9620$ &$0.0145$\\
\hline

\end{tabular}
 
 \caption{ Initial conditions and results for our models. The first set corresponds to Eq.~(\ref{case1a}) (using first set of parameteres of Table \ref{tkwz1}), the second to Eq.~(\ref{case2}) (using the first of Table  \ref{tkcc1}), the third to Eq.~(\ref{case3}) (using the second set of parameters of Table \ref{tkwz1}) and the fourth to Eq.~(\ref{case4}) (using the second set of parameteres of Table \ref{tkcc1}).  }
 \label{tbrans}
\end{table}

\begin{figure}[h!]
\centering
\includegraphics[width=1.1\textwidth]{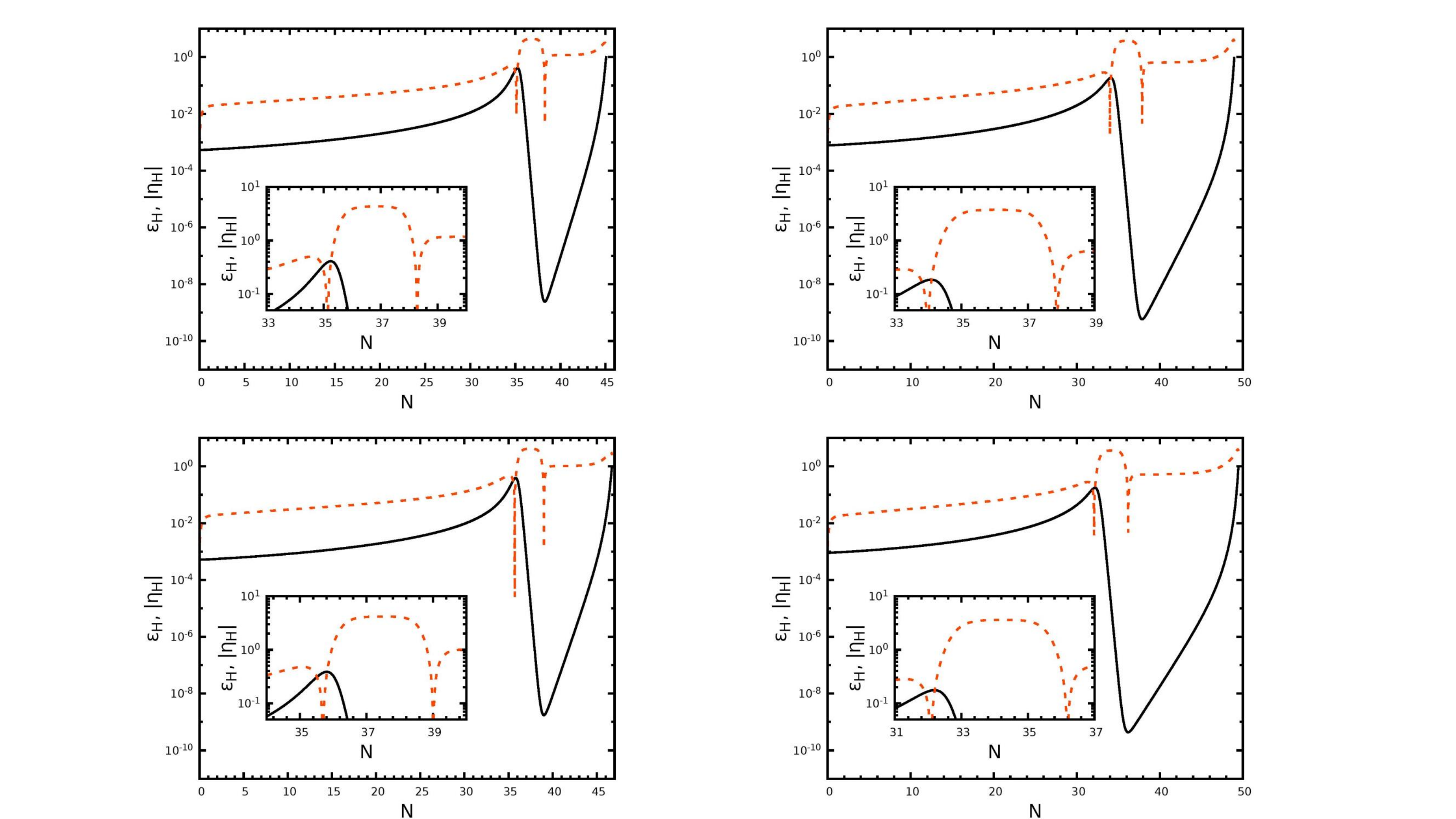}
\caption{Slow-roll parameters derived from Eq.~(\ref{22}). Details in the main text.}
\label{fbrahta}
\end{figure}

\par
As we mentioned before, an interesting feature which was be proposed in Ref.~\cite{Mishra19} is that a non modified model with spectra index $n_s$ bigger than the value of the prediction of Planck results can give us desirable values when it is modified with an inflection point. It is claimed that this feature emanates because the reinforcement near to the inflection point generates $\approx 10$ e-folds.
This is essential in order to construct models which are not in contradiction with the observable constrains.  
\par
After the evaluation of the Eq.~(\ref{21}), we can calculate the parameters $\varepsilon_H$ and $\eta_H$ using Eq.~(\ref{22}). All of our cases are depicted in Figure~\ref{fbrahta}, where Eqs.~(\ref{case1a}) corresponding  to up left panel, (\ref{case2}) to up right,  (\ref{case3}) to down left and (\ref{case4}) to down right.  One can proceed to an analytic evaluation of the power spectrum of cosmological perturbations, as it is given from the following equation \cite{Stewart:1993bc}:
\begin{equation}
 \label{24}
P_R \simeq \frac{1}{8 \pi^2} \frac{H^2}{\varepsilon_H}.
\end{equation}
which is valid within the slow-roll approximation. However, in Figure~\ref{fbrahta} we notice that this  approximation fails to be regarded as an accurate approximation because the parameters $\varepsilon_H$ and $\eta_H$ are not smaller enough than one. So the slow-roll regime cannot be fulfilled. Specifically, as it is depicted in Figure~\ref{fbrahta} the slow-roll parameter $\varepsilon_H$ is close to the value of one and $\eta_H$ is bigger than $3$.
There are many arguments (see Refs.~\cite{Ballesteros:2017fsr,Ozsoy:2018flq,Gao:2018pvq,Cicoli:2018asa,Dalianis:2018frf,Garcia-Bellido:2017mdw,Hertzberg:2017dkh,Mahbub:2019uhl,Mishra19}) in favor of solving the exact equations of field's perturbation in order to derive the power spectrum. 
Therefore, it is imperative to solve the exact perturbation system and to forsake the slow-roll approximation. We proceed with the numerical solution of field's perturbation.

\subsection{Numerical solution of field perturbation}
As it is advocated in the previous section, the slow-roll approximation fails to reproduce the proper power spectra and hence the masses and abundances of PBHs. 
Although, we can achieve having spikes within the slow-roll approximation using Eq.~(\ref{24}), the solution is not reliable for reasons described above and as a result we need to proceed to a more accurate evaluation. Many previous works in the literature about the production of PBH in single field inflation have highlighted the inaccuracy of the slow-roll approximation and the importance of evaluating the correct power spectrum. In this section we present the numerical equations of  perturbation, based on the analysis in Ref.~\cite{Ringeval:2007am}.
\par
The matter power spectrum can be evaluated by using the variance of fluctuation, which in homogeneous background is pictured by the curvature perturbation $R$. So the power spectrum is measured by:
\begin{equation}
\left<R_kR_{k ^ \prime} \right>= (2 \pi)^3 \delta^3(k+k ^ \prime)(2 \pi^2 /k^3)P_R(k).
\end{equation}
By solving the equations of perturbation, given below,  we can calculate the dimensionless power spectrum of comoving perturbation from:
\noindent
\noindent
\begin{equation}
\label{27}
P_R=\frac{k^3}{2 \pi^2} \left|R_k \right|^2,
\end{equation}
where $k$ is the comoving wave number of the Fourier mode and 
it is associated with the Mukhanov-Sassaki variable $Q_k$:
\begin{equation}
R_k=Q_k\frac{dN}{d \chi}.
\end{equation}
The Mukhanov-Sassaki variable then is given by:
\begin{equation}
Q_k= \delta \chi + \Psi \frac{d \chi}{dN},
\end{equation}
where $\delta \chi$ is the field perturbation and $\Psi$ the Bardeen potential.

The linear equations for the perturbation of the field, $\delta \chi$, in e-fold times reads: 
\begin{equation}
 \label{25} 
\frac{d^2 \delta\chi}{dN^2}=- \left(  3-\frac{1}{2} \left(\frac{d \chi}{d N}\right)^2 \right) \frac{d \delta \chi}{dN}-\frac{1}{H^2}\frac{d^2V}{d\chi^2} \delta \chi -\frac{k^2}{a^2 H^2} \delta \chi +4 \frac{d\Psi}{dN} \frac{d \chi}{dN}-\frac{2 \Psi}{H^2}\frac{dV}{d\chi}
\end{equation}
where the Bardeen potential $\Psi$ is considered by the equation:
\begin{equation}
 \label{25b}
\frac{d^2 \Psi}{dN^2}=-\left(  7-\frac{1}{2} \left(\frac{d \chi}{d N}\right)^2 \right) \frac{d \Psi}{d N}- \left( 2\frac{V}{H^2} +\frac{k^2}{a^2 H^2}\right)\Psi -\frac{1}{H^2}\frac{dV}{d\chi} \delta \chi.
\end{equation} 


\par
We need to solve numerically the Eqs.~(\ref{25}) and (\ref{25b}). The numerical procedure is based on Ref.~\cite{Ringeval:2007am}. In order to do that we need the initial conditions of field perturbation $\delta \chi$. The initial conditions are taken for subhorizon states, when $k\gg a H$, in order to be in the Bunch-Davies vacuum\cite{Bunch:1978yq}, where:
\begin{equation}
a Q_k \rightarrow \frac{e^{-ik\tau}}{\sqrt{2k}}.
\end{equation} 
Hence, these initial conditions for the perturbation are given:

\begin{equation}
\delta \chi_{ic}=\frac{1}{\sqrt{2 k}} \frac{1}{a_{ic}}, \quad\left(\frac{d \delta  \chi}{d N}\right)_{ic}= -\frac{1}{a_{ic}\sqrt{2k}}\left(1+i\frac{k}{a_{ic}H_{ic}}\right)
\label{26b}
\end{equation}
and for the Bardeen potential:
\begin{equation}
\begin{aligned}
\Psi_{ic}=\frac{1}{2 \left( \varepsilon_{H,ic} -\frac{k^2}{a_{ic}^2 H_{ic}^2} \right) } \left(  \left( \frac{d\chi}{dN}\right)_{ic} \left( \frac{d \delta\chi}{dN}\right)_{ic}+\delta \chi_{ic}\left[3\left(\frac{d\chi}{dN}\right)_{ic} +\frac{1}{H_{ic}^2}\left(\frac{dV}{d\chi}\right)_{ic}\right]\right)\\
\left( \frac{d \Psi}{dN}\right)_{ic}= \frac{1}{2} \left( \frac{d\chi}{dN}\right)_{ic} \delta \chi_{ic} -\Psi_{ic}\,,
\end{aligned}
\end{equation}
\noindent
where with the subscription $ic$, we denote the initial condition for each quantity. 
One can solve Eqs.(\ref{25}) and (\ref{25b}) twice for the real and imaginary parts for each $k$ mode as it is more convenient, but it is always possible to evaluate it in the full complex plane. For each $k$ of interest we solve the equation until the solution converges and we keep this value. After that we need to normalize the power spectrum, as its amplitude at $k_*=0.05Mpc^{-1}$ is well known.

\begin{figure}[h!]
\centering
\includegraphics[width=1.2\textwidth]{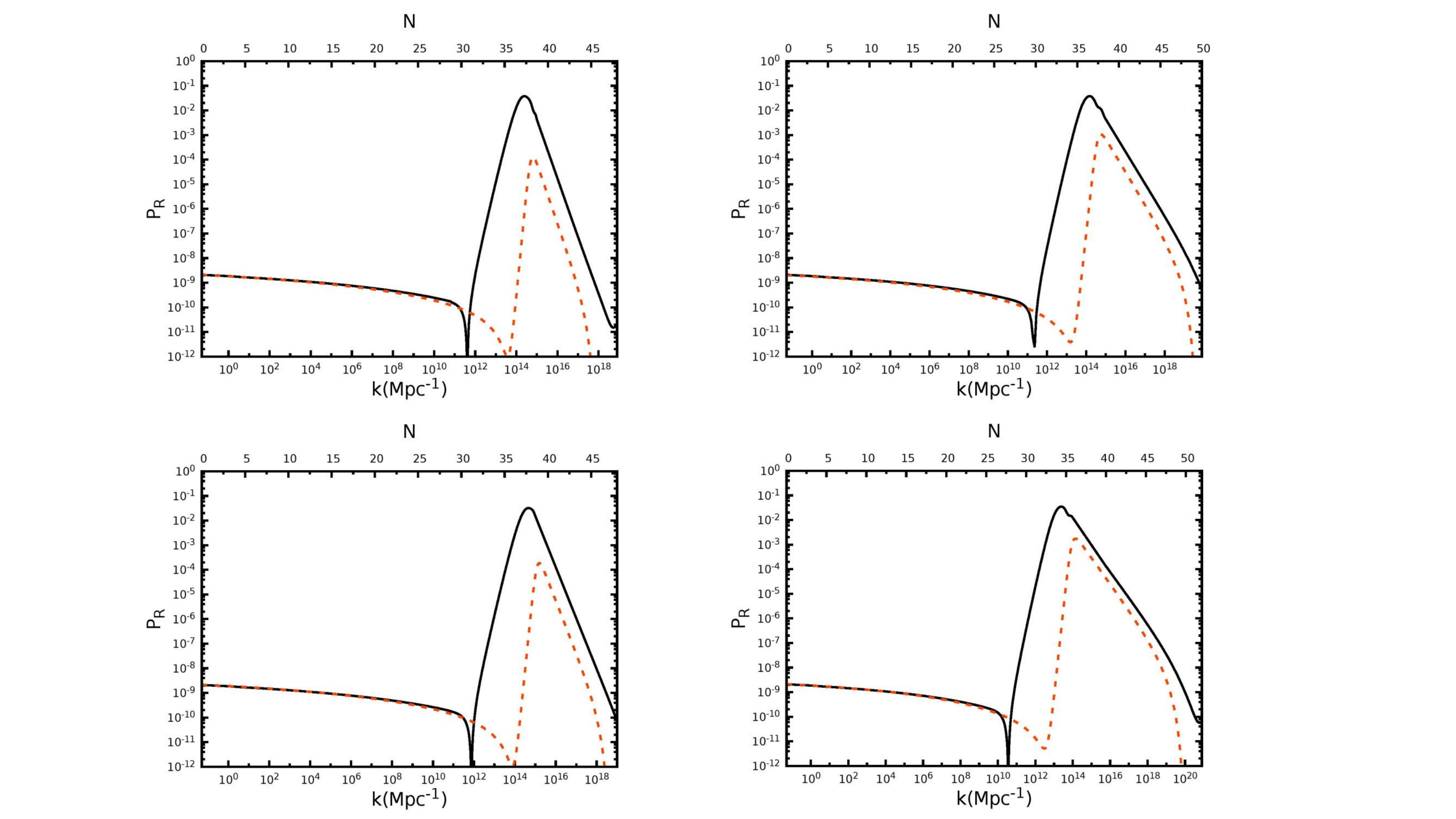}

\caption{ Power spectra by solving the equations of perturbation (black line) and via slow-roll approximation (dashed line).}
\label{fbrapr1}
\end{figure}
\par 
We solve numerically the  equations Eqs.~(\ref{25}) and (\ref{25b}) for superpotentials from Eqs.~(\ref{case1a}), (\ref{case2}), (\ref{case3}) and (\ref{case4}) in accordance with the set of parameters and initial conditions described above. Our results are depicted in Figure~\ref{fbrapr1} with the case  (\ref{case1a}) corresponding  to up left, (\ref{case2}) to up right,  (\ref{case3}) to down left and (\ref{case4}) to down right. We remark that we have a notable difference between the slow-roll approximation (dashed lines) given in Eq.~(\ref{24}) and the exact equations of perturbation (black lines). This conclusion coincides with those of previous works, which claim that the numerical solution for perturbation is necessary in order to have the valid power spectrum. Although the slow-roll approximation gives us essential enhancement in the power spectrum, it fails to reproduce  the peak height. This failure can be explained by noticing that the absolute value of $\eta_H$ becomes larger than $3$ around the region where PBHs are generated. To sum up, Eq.~(\ref{24}) is not regarded as a good evaluation for the power spectrum and the deduction of its failure is envisaged in the figures of this section.
   
%

\subsection{PBH Production}
  
\par
Up until now we developed a mechanism in order to generate significant peaks in the power spectrum. We interpret those fluctuation modes to the formation of PBHs. Specifically, when an intense perturbation, produced during the stage of inflation, reenters the Hubble radius, $k=a H$, it can form a PBH. In this section we evaluate the masses of PBHs, $M_{PBH}$, and their abundances, $\Omega_ {PBH} / \Omega_ {DM}$, for our models. The PBH mass depends on the comoving wave number $k$ and the abundance depends on the PBH mass and on the power spectrum, which both of them depend on $k$. The calculation described below is done by using both the PS approach of gravitational collapse \cite{Press74} and the PT one\cite{Bardeen:1985tr,Young:2014ana}. 
 
\par 
As we mentioned PBHs are produced when scales start reentering the horizon after the cessation of inflationary expansion. Their mass is associated with the mass inside the Hubble horizon in the straightforward form:
\begin{equation}
M_{PBH}= \gamma M_H=\gamma \frac{4}{3} \pi \rho H^{-3}
\label{43a1}
\end{equation}  
\noindent
where $\rho$ is the energy density of the Universe during the collapse and $\gamma$ is the correction factor which depends on the gravitational collapse. From Eq.~(\ref{43a1}) we can obtain a ratio of $M_{PBH}$ and the mass at the horizon, denoted with subscription $eq$, in order to have a relation between the $M_{PBH}$ and the comoving wave number. Hence by the conservation of entropy $d(g_s(T)T^3 a^3) /dt =0$ (where $T$ denotes the temperature), we find that:
\begin{equation}
M_{PBH}= \gamma M_{H(eq)}\left (\frac{g(T_f)}{g(T_{eq})} \right )^{1/2} \left (\frac{g_s(T_f)}{g_s(T_{eq})} \right )^{-2/3} \left( \frac{k}{k_{eq}}\right)^{-2},
\label{43a2}
\end{equation} 
\noindent
where $\rho =g(T)T^4$, $g(T)$ is the effective number of degrees of freedom in the radiation density and $g_s(T)$ is the same in entropy density. Moreover, $g(T_f)$ is the effective number of degrees of freedom when the PBHs are produced. 
Supposing that $g(T)=g_s(T)$, we compute the mass as a function of $k$ \cite{Ballesteros:2017fsr}:
\begin{equation}
\label{43}
M_{PBH}=10^{18}  \left(\frac{\gamma}{0.2}\right)  \left(\frac{g(T_f)}{106.75}\right)^{-1/6} \left(\frac{k}{7 \times 10^{13} Mpc^{-1}  }\right)^{-2} \textsl{g}.
\end{equation}

The variance of curvature perturbation $\sigma$ is related to the power spectrum by the following expression:
\begin{equation}
\label{40}
\sigma^2 (M_{PBH}(k))= \frac{4(1+\omega)^2}{(5+3\omega)^2}  \int \frac{dk' }{k'} \left(\frac{k'}{k}\right)^4 P_R(k') \tilde W^2\left(\frac{k'}{k}\right)
\end{equation}
\noindent
where $\omega$ is the equation of state $\omega= p/ \rho$  and in radiation dominated epoch is equal to $1/3$. $\tilde W \left(\frac{k'}{k}\right)$ is the Fourier transform of the window function. Throughout this work we will use the Gaussian distribution $\tilde W(x)=e^{ -x^2/2}$ for this  function. The density power spectrum is defined as:
\begin{equation}
P_{\Delta}=\frac{4(1+\omega)^2}{(5+3\omega)^2} \left(\frac{k'}{k}\right)^4P_R(k')
\end{equation}

As the evaluation of the $\sigma$ is the same for both of these method, we proceed to the calculation of the fraction of the Universe collapsing in PBH of mass $M_{PBH}$,  $\beta$. In PS approach the mass fraction $\beta_{PS}$ is given by the probability that the overdensity $\delta$ is above a certain threshold of collapse, denoted as $\delta_c$. The fraction $\beta_{PS}$ for this approach reads as:
\begin{equation}
\label{42}
\beta_{PS}(M_{PBH})= \frac{1}{\sqrt{2 \pi \sigma ^2 (M)}} \int^{\infty}_{\delta_c} d\delta ~ e^{\frac{-\delta ^2}{2 \sigma^2(M)}} =\frac{1}{2} \text{Erfc}\left(\frac{\delta_c}{\sqrt{2 }\sigma}\right)
\end{equation}
\noindent
where $\text{Erfc}(x)$ is the complementary error function:
\begin{center}
$\text{Erfc}(x) = 1-\frac{\text{sgn}(x)}{\sqrt{\pi}} \gamma (\frac{1}{2},x^2).$
\end{center}
\noindent
The Eq.~(\ref{42}) is computed using the incomplete Gamma function:
\begin{equation}
\beta_{PS}(M_{PBH})=\frac{\Gamma\left(\frac{1}{2}, \frac{\delta_c^2}{2 \sigma^2}\right)}{2\sqrt{\pi}}. 
\label{42b}
\end{equation}
On the other hand, in PT one needs to calculate the number density of peaks above a threshold given:
\begin{equation}
n_{peaks}= \frac{1}{(2 \pi)^2} \left(  \frac{\left\langle {{k^2}} \right\rangle}{3}\right)^{3/2}\left( \left(\frac{\delta_c}{\sigma}\right)^2-1\right)exp\left(  - \frac{\delta_c^2}{2 \sigma^2}\right)
\end{equation}
where $\left\langle {{k^2}} \right\rangle$ reads:
\begin{equation}
\left\langle {{k^2}} \right\rangle= \frac{1}{\sigma^2} \int^{\infty}_{0} \frac{dk'}{k'} {k'}^2\tilde W \left(\frac{k'}{k}\right)^2 P_{\Delta}(k')
\end{equation}
The mass fraction for a flat Universe  assuming a  Gaussian  distribution is given by $\beta_{PT}=n_{peaks}(2\pi)^{3/2}(1/k)^3$ \cite{Green:2004wb}.

Finally, the present abundance of PBH is given by the integral:
\begin{equation}
f_{PBH} = \int d\ln M \frac{\Omega_ {PBH}}{\Omega_ {DM}}
\end{equation}
\noindent
where
\begin{equation}
\label{44}
\frac{\Omega_ {PBH}}{\Omega_ {DM}}= \frac{\beta(M(k))}{8 \times 10^{-16}} \left(\frac{\gamma}{0.2}\right)^{3/2} \left(\frac{g(T_f)}{106.75}\right)^{-1/4}\left(\frac{M(k)}{10^{-18} \textsl{g}}\right)^{-1/2}.
\end{equation}
For the values of $\gamma$ and  $g(T_f)$, we choose $\gamma =0.2$ \cite{Carr1975} and $g(T_f)=106.75$.  We take into consideration that the recent analyses during radiation domination, which suggest that $\delta_c \approx 0.4-0.5$ \cite{Harada:2013epa,Musco:2008hv,Musco:2004ak,Musco:2012au,Musco:2018rwt,Escriva:2019phb,Escriva:2020tak}. 
\begin{figure}[h!]
\centering
\includegraphics[width=1.2\textwidth]{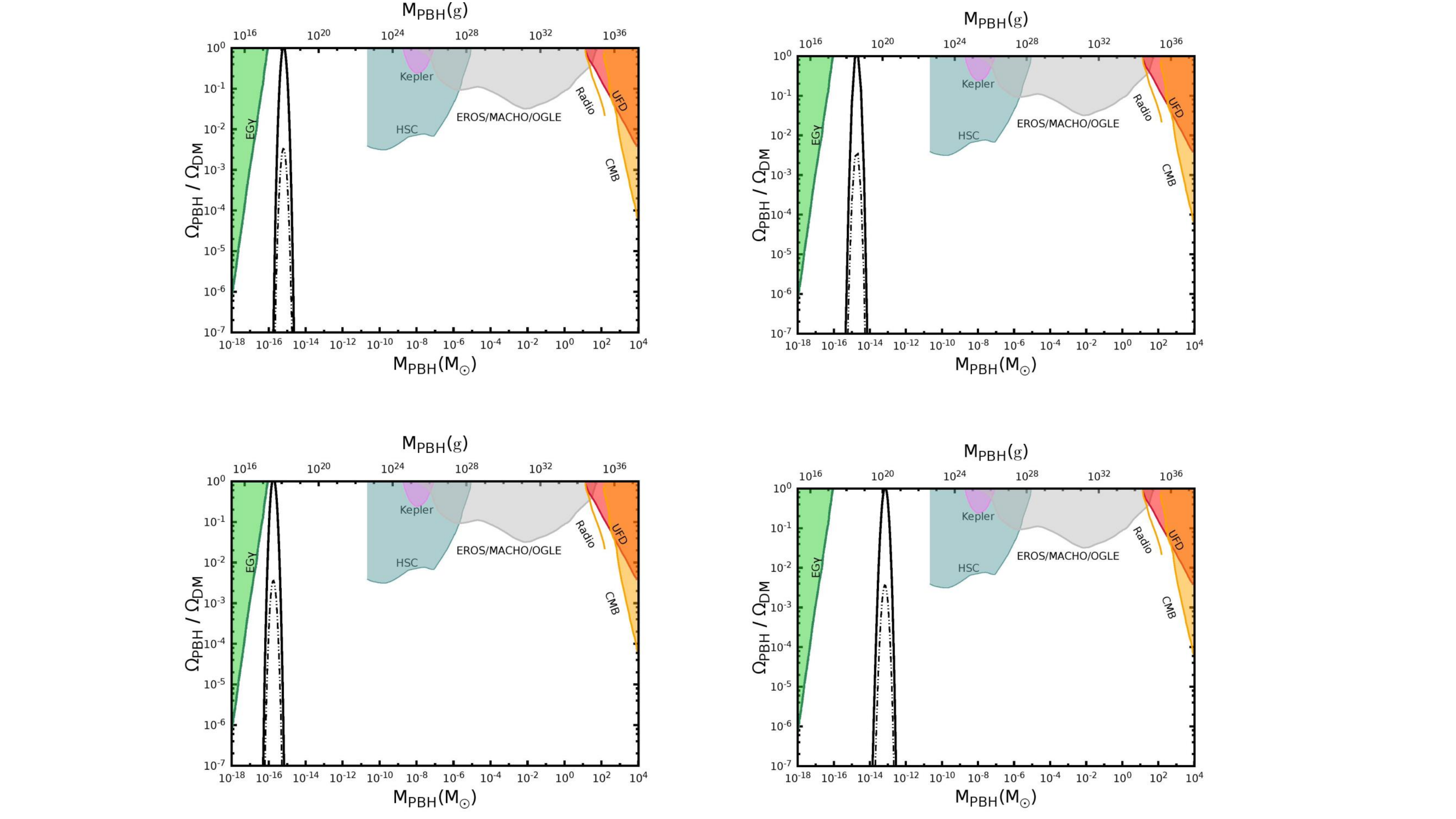}
\caption{The fractional abundance of PBH  using PS approach and PT and the bounds given by \cite{Carr:2009jm,Inoue:2017csr,Montero-Camacho:2019jte,Katz:2018zrn,Poulin:2017bwe,Capela:2013yf,Niikura:2017zjd,Wyrzykowski:2011tr,Griest:2013esa,Tisserand:2006zx,Ali-Haimoud:2016mbv,Gaggero:2016dpq}. These observation data are from extragalactic gamma ray from PBH evaporation ($EG\gamma$) \cite{Carr:2009jm},  microlensing for Subaru (HSC), Eros/Macho/Ogle \cite{Capela:2013yf,Niikura:2017zjd,Wyrzykowski:2011tr} and Kepler\cite{Griest:2013esa} dynamical heating of ultra faint dwarf (UFD) \cite{Tisserand:2006zx}, CMB measurements \cite{Ali-Haimoud:2016mbv} and radio observation \cite{Gaggero:2016dpq}. Details are found in the main text.
 }
\label{fbraabu1}
\end{figure}

\begin{table}[h!]

\centering
 \begin{tabular}{||c c c c c c||} 
 
 \hline
  & $P^{peak}_{R}$& $M^{peak}_{PBH}(M_\odot)$  &$f_{PBH}^{PS} $ & $f_{PBH}^{PT} $ & $\delta_c$  \\ [0.5ex] 
 \hline\hline
1.& $3.77\times 10^{-2}$ & $6.12 \times 10^{-16}$&$2.02\times 10^{-3}$ & $ 0.682$ & $0.48$   \\
\hline
2. &$3.79 \times 10^{-2}$ & $2.18 \times 10^{-15}$& $2.72\times 10^{-4}$&$  0.889$ & $0.48$   \\ 
 \hline
3.& $3.21 \times 10^{-2}$ & $1.84 \times 10^{-16}$&$2.20\times 10^{-3}$& $ 0.761$ & $0.45$   \\
\hline
4. &$3.50 \times 10^{-2}$ & $6.60\times 10^{-14}$& $2.40\times 10^{-3}$ &$  0.734$ &$0.45$   \\ 
\hline

 \end{tabular}
  \caption{ The values of the peak of power spectrum, their fractional abundance and th $\delta_c$ for the superpotentials (\ref{case1a}),    (\ref{case2}), (\ref{case3}) and (\ref{case4}) respectively, with initial conditions shown in Table~\ref{tbrans}.}
 \label{tbraomega}
\end{table}

\par
With this formalism we can calculate the fraction abundances for all cases and we get the desirable abundances in order to explain the origin of the dark matter. Our results are illustrated in Figure~\ref{fbraabu1}  with Eqs.~(\ref{case1a}) corresponding  to up left, (\ref{case2}) to up right,  (\ref{case3}) to down left and (\ref{case4})to down right. With solid lines we present the outcome for the PT, whereas with dashed lines for the PS approach. We observe a difference between these two methods, as it is previously shown in Ref.~\cite{Young:2014ana}. However, this difference can be compensated by the fact that the result is sensitive to the value of $\delta_c$. This means that, by a slight change of $\delta_c$, the fractional abundances evaluated by PS can give results comparable to PT. We sum up our results in Table~\ref{tbraomega}, where one can notice this difference for a given $\delta_c$. 
\section{PBH by modifying the K\"ahler potential}
\label{case1}
\par 
So far we study potentials which can generate PBH that might account for a significant fraction of the dark matter in the Universe. In the previous section we modify well known superpotentials derived by SU(2,1)/SU(2)$\times$U(1) symmetry in order to produce such peaks. In this section we break this symmetric by adding extra terms in the K\"ahler potential and then we reduce the problem to single field inflation. Hence we present a different way to generate PBH by introducing difference in K\"ahler potential.
\par
In a previous work it was shown that PBH can be generated through the scheme \cite{Nanopoulos:2020nnh}:

\begin{gather}
K=-3 \ln( T+ \bar{T} - \frac{\varphi \bar{\varphi}}{3}+ ce^{-b(\varphi +\bar{\varphi})^2}(\varphi +\bar{\varphi})^4)\\
W= \frac{\hat \mu}{2} \varphi^2  - \frac{\lambda}{3} \varphi^3
\end{gather} 
where $c$ and $b$ are parameters. In Ref.~\cite{Nanopoulos:2020nnh} the superpotential was given by the Wess-Zumino form, while in this work we consider that the superpotential is given by the Cecotti form. Specifically, we have:
\begin{equation}
K=-3 \ln( T+ \bar{T} - \frac{\varphi \bar{\varphi}}{3}+ F( T+ \bar{T},\varphi+ \bar{\varphi}))
\label{ciii1}
\end{equation} 

\begin{equation}
W=\sqrt{3}m \varphi\left(T- \frac{1}{2}\right)
\label{ciii2}
\end{equation}
\noindent
where $F( T+ \bar{T},\varphi+ \bar{\varphi})$ is a function of both chiral fields. Choosing:
\begin{equation}
\label{ciii3}
 F( T+ \bar{T},\varphi+ \bar{\varphi})=e^{-d (T+ \bar{T})}\left(c_4 e^{-b_3(T+ \bar{T})}(T+ \bar{T})^2 +\bar{\lambda}(\varphi+ \bar{\varphi}+ \bar{\mu})\right)
\end{equation}
\noindent
where $b_3$, $c_4$, $\bar{\lambda}$, $\bar{\mu}$  and $d$ are constants, we can evaluate the kinetic term and the scalar potential, using Eqs. (\ref{02}) and (\ref{3}). 

\par
The Starobinsky effective scalar potential arises when $c_4=\bar{\lambda}=0$ while taking into account that the real part of $T$ is the inflaton field and $\varphi$ is the modulo field. The fields $\varphi$ and $\Im T$ will be forced to vanish during inflation \cite{Kallosh:2010xz}. Because the desirable inflationary direction occurs at $\varphi=\Im T=0$, we have to show that it can always be stabilized. So, we need to construct the mass squared matrix (\ref{massmatrix}) and check that the fields, which will be discarded, are not tachyons. 
As in the previous section, by using the methodology presented in Refs.~\cite{Ellis:2019bmm,Ellis:2018xdr} we can prove that the field $\varphi$ and $T$ are stabilized in both real and imaginary direction.  We deduce that for our set of parameters the masses for both real and imaginary fields, fixed at $\left\langle {{T}} \right\rangle= \frac{1}{2}$ and  $\left\langle {{\varphi}} \right\rangle=0$ are always positive. We remark that for our set of parameters we do not need additional terms in order to achieve stabilization along the inflationary direction.

 \par
 We proceed to the calculation of the effective scalar field in the inflationary direction. From Eq. (\ref{02}) we find the K\"ahler metric. In general, the first component of this metric, which involves the inflaton field T is given by:

\begin{equation}
{K_{\bar{T}T}}=\frac{3 \left(\left( F^{\prime}+1\right)^2-
   F^{\prime\prime} \left( F+\bar{T}-\frac{\varphi  \bar{\varphi }}{3}+T\right)\right)}{\left( F+\bar{T}-\frac{\varphi  \bar{\varphi }}{3}+T\right)^2}
\label{7}
\end{equation} 
where with prime we denote the derivatives in respect to chiral field $T$.
 We study the case $\varphi= \Im T=0$, $\Re T=\phi$ and $F=F(T+ \bar{T},\varphi+ \bar{\varphi}) \neq 0$. So, Eq.~({\ref{7}) is simplified and, because we must have canonical kinetic term in the Langrangian, we have the following redefinition of the field:
 \begin{center}
  $\frac{1}{2}\partial_{\mu}\chi \partial^{\mu}\chi= K_{\phi \phi}\partial_{\mu}\phi \partial^{\mu}\phi \Rightarrow  $
 \end{center}
 
 \begin{equation}
\label{9case3}
\frac{d \chi}{d \phi}= \sqrt{2 K_{\phi \phi }} .
\end{equation}

\par
The effective scalar field in this case is computed numerically by using the redefinition of the field, Eq.~(\ref{9case3}), and the Eq.~(\ref{3}). We conclude that we can achieve scalar potentials with canonical kinetic terms that have the recommendable inflection point in order to lead to the PBH formation. In Table ~\ref{tabucase3a} we present some parameters and in Figure~\ref{f1b} we plot the first set of them. The parameter $m$ is fixed in order to satisfy the constraint of power spectrum at CMB scales. The role of the parameter $d$ is to fix the observable constraints, because the modification we put generates extra e-folds which lead the $\chi_*$, when $k_*=aH_*$, closer to $\chi_{end}$, when inflation ends, as it is observed in Ref.~ \cite{Mishra19}. The parameter $b_3$, which needs a lot of fine-tuning, gives us the peak height. The other parameters are needed in order to achieve the shape of the inflation point.  In the next section we will discuss the choice of parameter in a more detailed way.

Taking everything into account, the effective scalar potential has a region for the reheating, the desirable inflection point and moreover it is flat for large values of $\chi$. We notice that, in contradiction with the previous section, we have a potential, which is flat at large values of the field and it satisfies the CMB constraints, as we will see later. 

\begin{table}[h!]
\centering
 \begin{tabular}{|c|c|c|c|}
\hline
 & $c_4$ & $b_3$ & $d$\\
\hline
1 & $9$ &$7.6626159$ &$0.048$  \\
\hline
2 & $8.8$ &$7.5151043$ &$0.054$\\

\hline

\end{tabular}
 
 \caption{Parameters for the potential derived from \ref{ciii1} and \ref{ciii2}. We take  $\bar{\lambda}=-0.5$ and $\bar{\mu}=1$.}
 \label{tabucase3a}
\end{table}

\begin{figure}[h!]
\centering
\includegraphics[width=80mm]{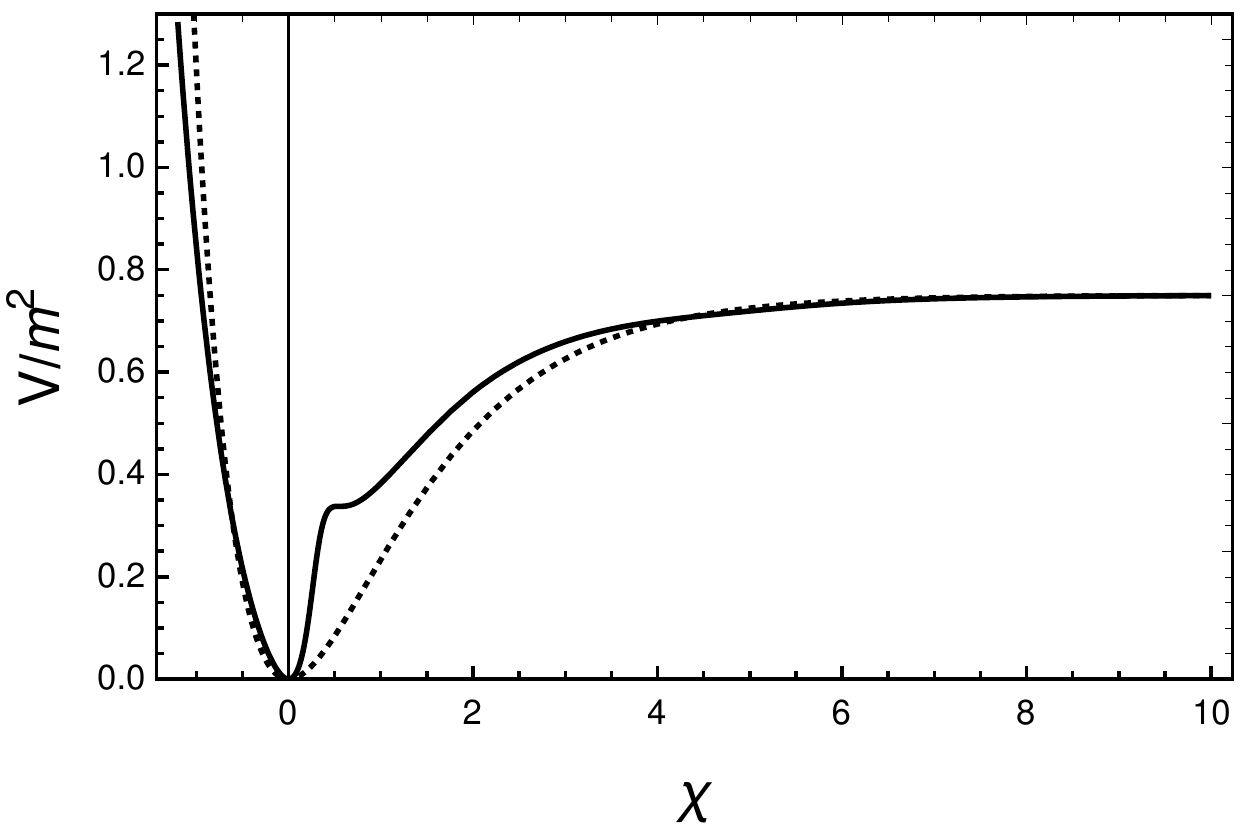} 
\caption{ Potential as given from Eq.~(\ref{3}) using first set of parameters  of the Table~\ref{tabucase3a} (solid). In the case where $c_4$ and $\bar{\lambda}$ are zero (dotted), we recover the Starobinsky potential. We notice an inflection point near $\chi=0.5$.}
\label{f1b}
\end{figure}


We calculate the evolution of the field and the slow-roll parameters. This case is more labored than the previous one, where we have modified superpotentials, because one should always take into consideration the redefinition of the field which is not given by an analytical expression anymore. Hence the background equation (\ref{21}) must be computed simultaneously with the Eq.~(\ref{9case3}) until the inflation ends, $\varepsilon_H = 1$ is satisfied. The number of e-folds is determined as before, as the duration between the two Hubble crossing conditions. 

 We apply the formalism described in the previous section. For initial condition we use the values in Table \ref{tabucase3a}. The initial condition for the field with canonical kinetic term $\chi_0$ is evaluated if we just know the corresponding field $\phi_0$, as seen in Eq.~(\ref{9case3}).

The initial condition for the velocity is given by the slow-roll attractor, as in the previous section. The values of $\phi_0$ are selected in order not only to be in the range when the potential is almost flat, but also to give the correct prediction for $n_s$ and $r$ (see Eq.~(\ref{21a})). We need to remark here that the initial conditions in all of our cases remain to be in the region, when the potential is regarded to be in CMB scale.

The slow-roll parameters $\varepsilon_H$ and $\eta_H$ are illustrated in Figure ~\ref{f4}. By noticing all plots of slow-roll parameters in this paper, we can observe that the parameter $\varepsilon_H$ is closer to the value 1 in the left panels of Figure~\ref{fbrahta} than the right panel of this figure, as well as the Figure~\ref{f4}. One can observe that in cases which $\varepsilon_H$ is closer to 1, the inflation ends sooner.

In Figure~\ref{f4} one can observe that the validity of slow-roll approximation cannot be verified, as before. Specifically, we expect that the slow-roll approximation does not give the correct power spectrum, because the values of $\varepsilon_H$ are not small enough to one and the value of $\eta_H$ is bigger than 3. Consequently, we proceed again with solving the exact equations of perturbation.

\begin{table}[h!]
\centering
 \begin{tabular}{||c c c c||} 
 
 \hline
  &$\phi_0$  &$n_s$ & $r$ \\ [0.5ex] 
 \hline\hline
 1.&$11.2 $  &$0.9603$ & $0.00775$ \\ 

 \hline
 2.&$10.5$ &$0.9601$ & $0.00929$\\ 

 \hline
 
 \end{tabular}
 \caption{ Initial conditions of the fields,  $n_s$ and $r$ for the modified K\"ahler with corresponding to the entries of the Table~\ref{tabucase3a}.}
 \label{tabucase3b}
\end{table}
\noindent

\begin{figure}[h!]
\centering
\includegraphics[width=130mm]{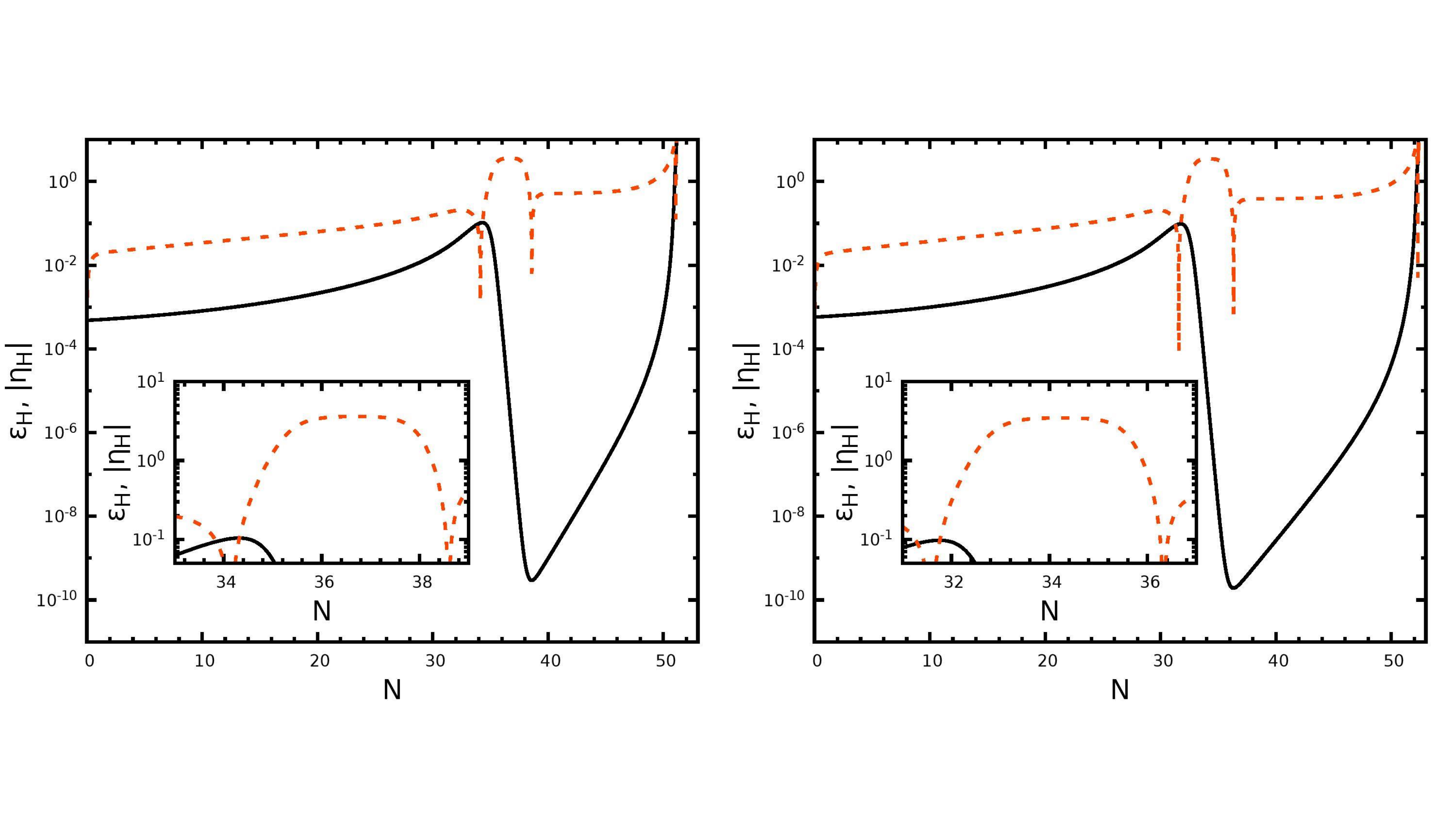}

\caption{Slow-roll parameters $\varepsilon_H$ (black lines), $\eta_H$ (dashed lines)  which corresponds to the first set first set (left panel) and second set (right panel) of parameters from the Tables~\ref{tabucase3a} and \ref{tabucase3b}.}
\label{f4}
\end{figure}


\par
We evaluate numerically the Eqs.~(\ref{25}) and (\ref{25b}) and we calculate the power spectrum from Eq.~(\ref{27}) using the methodology described previously. The power spectra derived by the numerical equations for curvature perturbation is depicted in Figure~\ref{fpr3a}. In the same figure we plot the power spectrum in the slow-roll approximation (\ref{24}). We can notice that the approximation fails to give the exact value of power spectrum, as it is underestimates its peak's height. In accordance with other works find that the slow-roll approximation is violated and this claim is shown by Figure~\ref{fpr3a}. Despite the value of the peak, it also fails to give the correct position of the peak.
\par
Observing in all plots of power spectrum computed in this work in Figure~\ref{fbrapr1} and \ref{fpr3a}, we can conclude that the value of peak's height depends on how close to one the slow-roll parameter $\varepsilon_H$ becomes. In this example, where we modify K\"ahler potential, the $\varepsilon_H$ is not close enough to one and as a result the peak's height underestimate one order of magnitude. In contradiction to this, one can notice in Figure~\ref{fbrapr1} that the left panels give an enhancement of more than two orders of magnitude.
 
\begin{figure}[h!]
\centering
\includegraphics[width=150mm]{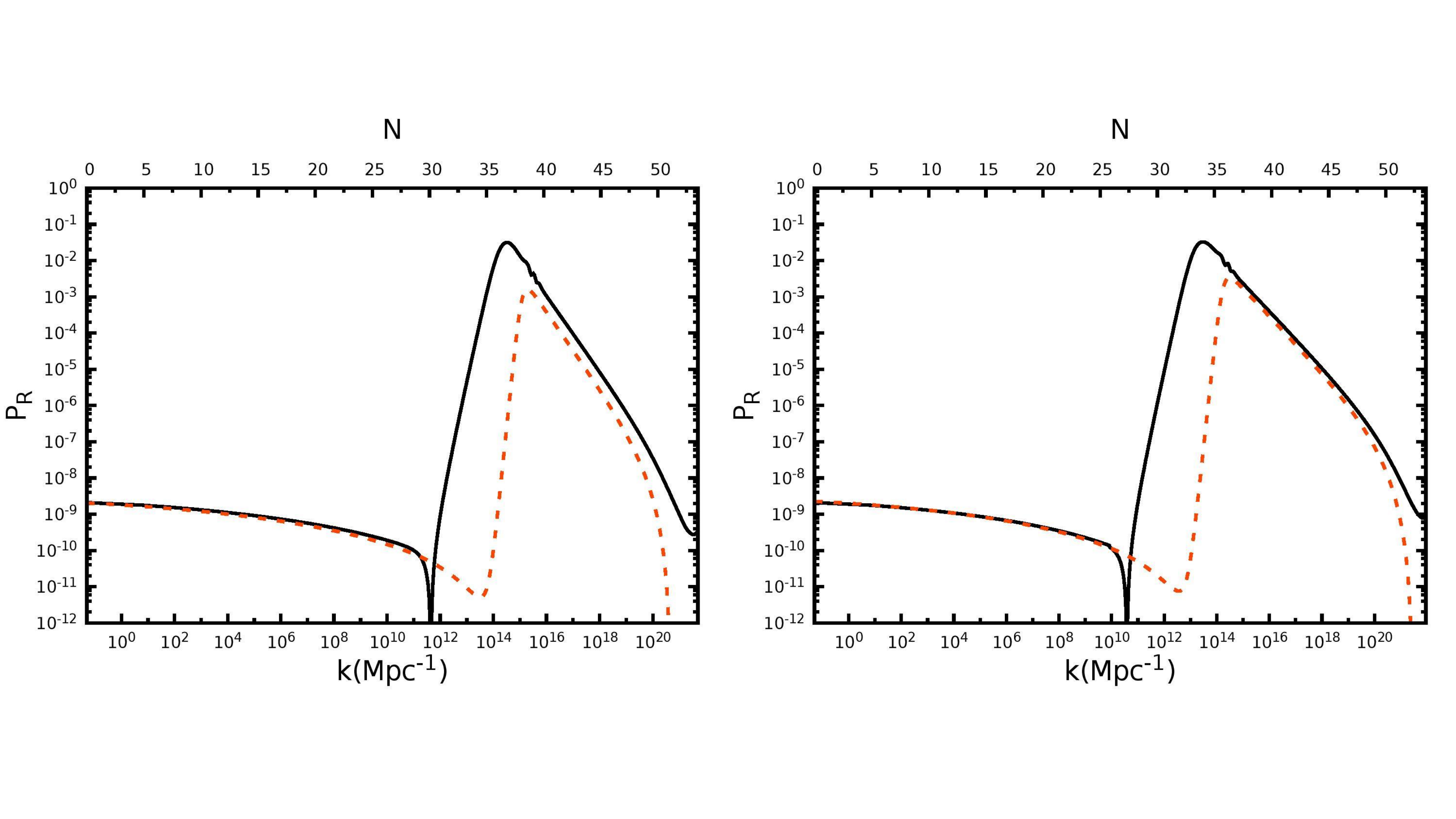}
\caption{ Power spectrum by solving numerically the linear perturbation (black line) and via slow-roll approximation (dashed line) for  case where K\"ahler potential is modified, first set (left panel) and second set (right panel) of parameters from the Tables~\ref{tabucase3a} and \ref{tabucase3b}.}
\label{fpr3a}
\end{figure}


By applying the methodology presented in the previous section we produce PBHs and we plot the fractional abundances in Figure \ref{f21}. We compute the fractional abundances by using both the PS approach and  the PT one. Our results for all sets of parameters are presented in Table~\ref{tabu43}. As before the PT gives a systematically higher abundance of PBH than the PS for a given threshold.

As we have completed two different ways which give us the production of PBH, one in which we modify superpotential and one in which we modify K\"ahler potential we can sum up some general aspects about the masses of PBHs.
The masses of PBH have a bound from below due to Hawking radiation \cite{Hawking1975}. On the other hand there is an upper restriction comes from nonobservation of PBH accretion on CMB \cite{Ricotti:2007au}. Our results are consistent with the constraints calculated in Ref.~\cite{Laha:2020ivk}. Finally, the mass fraction is sensitive to the critical density of collapse $\delta_c$ \cite{Ballesteros:2017fsr}. This issue will be analyzed further in the next section.

\begin{figure}[h!]
\centering
\includegraphics[width=150mm]{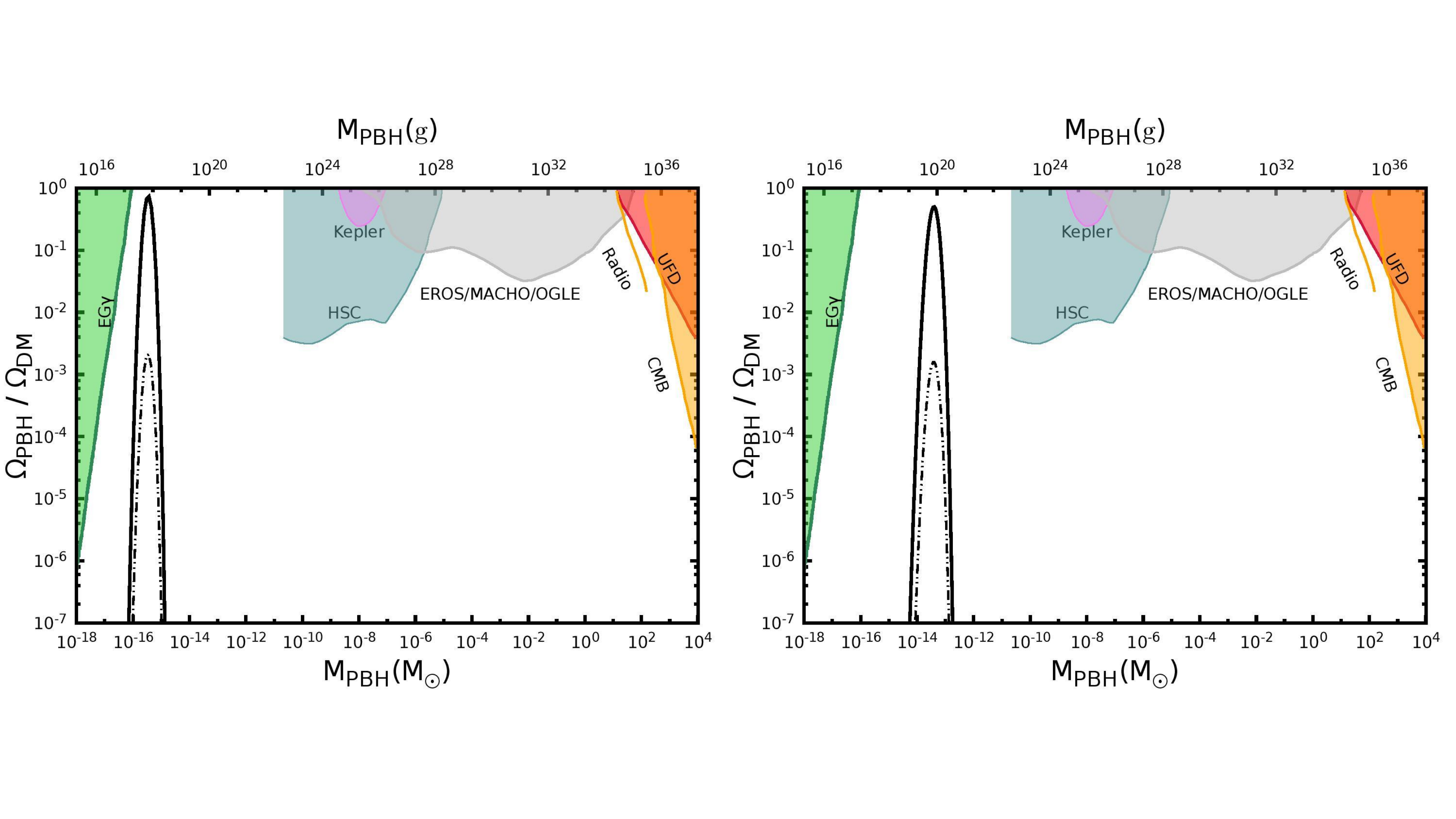}
\caption{The fractional abundance of PBH for the first set (left panel) and second set (right panel) of parameters in the Table~\ref{tabucase3a} and \ref{tabucase3b}, with observational constraints given in Figure \ref{fbraabu1}. Solid lines corresponds to the PT and dashed lines to the PS approach.}
\label{f21}
\end{figure}

\begin{table}[h!]

\centering
 \begin{tabular}{||c c c c c c||} 
 
 \hline
  & $P^{peak}_{R}$& $M^{peak}_{PBH}(M_\odot)$  &$f_{PBH}^{PS} $ & $f_{PBH}^{PT} $ & $\delta_c$  \\ [0.5ex] 
 \hline\hline
1.& $3.17 \times 10^{-2}$ & $3.66\times 10^{-16}$& $1.50\times 10^{-3}$ & $ 0.522$ & $0.45$   \\
 \hline
2.& $3.27 \times 10^{-2}$ &$ 3.67\times 10^{-14}$& $1.27 \times 10^{-3}$&$0.406$ & $0.45$   \\ 
 \hline
 \end{tabular}
  \caption{ The values of the peak of power spectrum for modified K\"ahler potential and their fractional abundance with correspondence to Tables~\ref{tabucase3a} and \ref{tabucase3b}.}
 \label{tabu43}
\end{table}


\section{Comments on the fine-tuning}

It is known that the formation of PBH by inflation demands fine-tuning. Many previous works have shown that, in order to achieve significant peaks in power spectrum, fine-tuning \cite{Hertzberg:2017dkh,Mahbub:2019uhl} is required. Even if one achieves to have an inflection point fine-tuning is needed in order to retrieve the proper enhancement in curvature power spectrum. In a previous work \cite{Hertzberg:2017dkh} the required fine-tuning has quantified in a way to approach the observable constraints. Here, we study this issue in our work by explaining how parameters in our potentials arise. Specifically, we present the case (\ref{case2}) with parameters presented in Table~\ref{tkcc1}  and we remark that the other set of parameters in the rest of cases come in the same way.

\begin{figure}[h!]
\centering
\includegraphics[width=150mm]{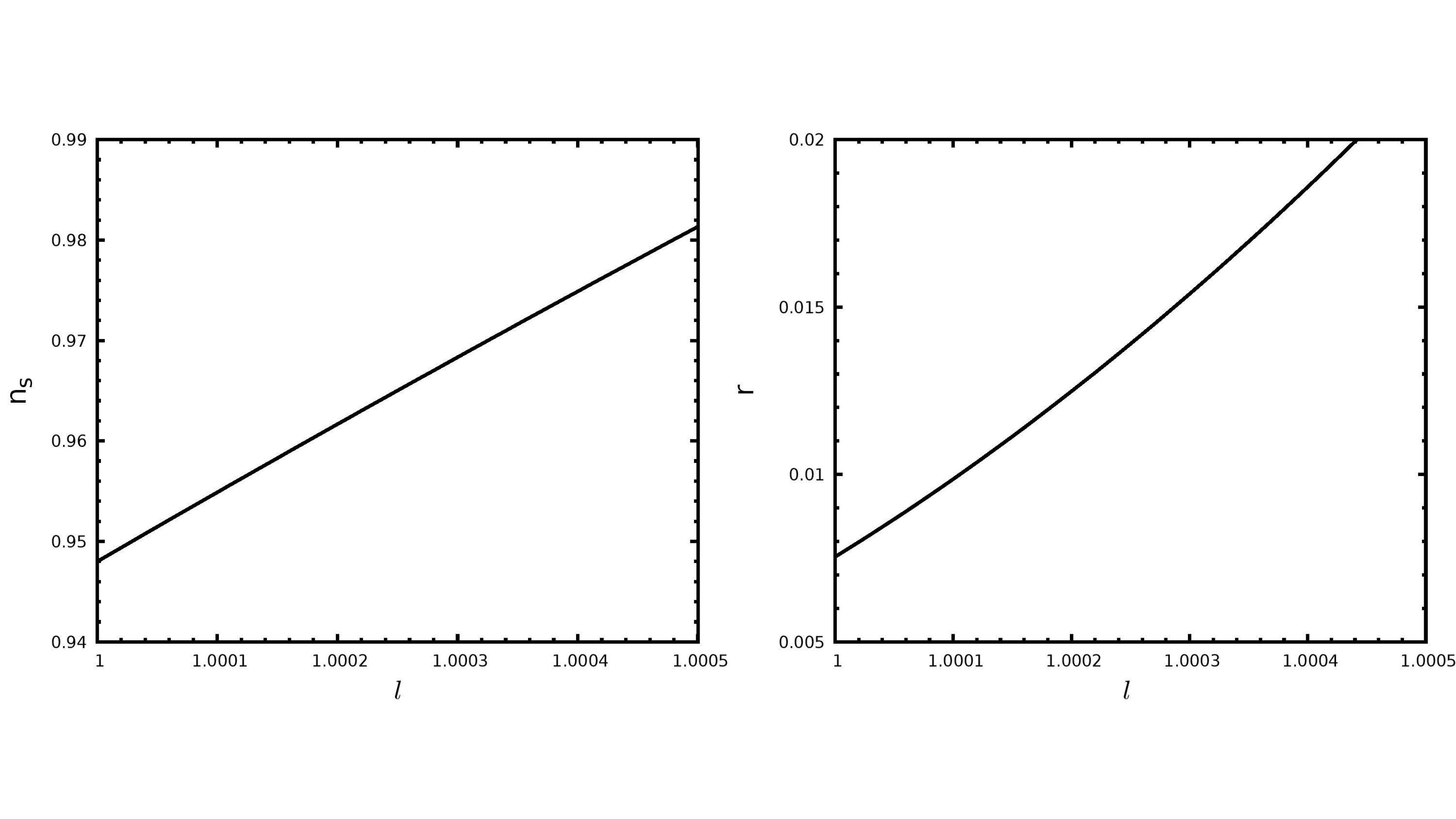}
\caption{How parameters arise in order to explain the observable constraints $n_s$ and $r$.  }
\label{fdns}
\end{figure}

So far, we always have a perturbation of a well-known theory by an exponential function and a polynomial. The fine-tuning in our work emanates from the cosmological constraints. First, the parameters $m$ and $\mu$ are set in order to be consequent of the CMB scales. Hence their values come from the constraints of power spectrum at $N=0$ which should have the value $2.1$ $\times$ $10^{-9}$. Moreover, we remark that the values of parameters  $c_2$ in  Table~\ref{tkwz1}, the parameter $l$ in Table~\ref{tkcc1} and $d$ in Table~\ref{tabucase3a} come from the restriction of spectral index $n_s$ and the ratio $r$ in order to be in accordance with Planck constraints. For instance, in Figure~\ref{fdns} we can  observe how we find the value of $l$ for the K\"ahler potential and superpotential given in Eq.~(\ref{case2}).  
We remark that parameters, such as $m$ and $\mu$, which determine the power spectrum at CMB scale and parameters, such as $d$, $c_2$  and $l$, which  determine $n_s$ and $r$ have a mild dependence of the others parameters described bellow. In Figure~\ref{fdns} we take into consideration the parameters by the first set from Table~\ref{tkcc1}. Same predictions of  $n_s$ and $r$ are derived by the other set of parameters too.

\begin{figure}[h!]
\centering
\includegraphics[width=150mm]{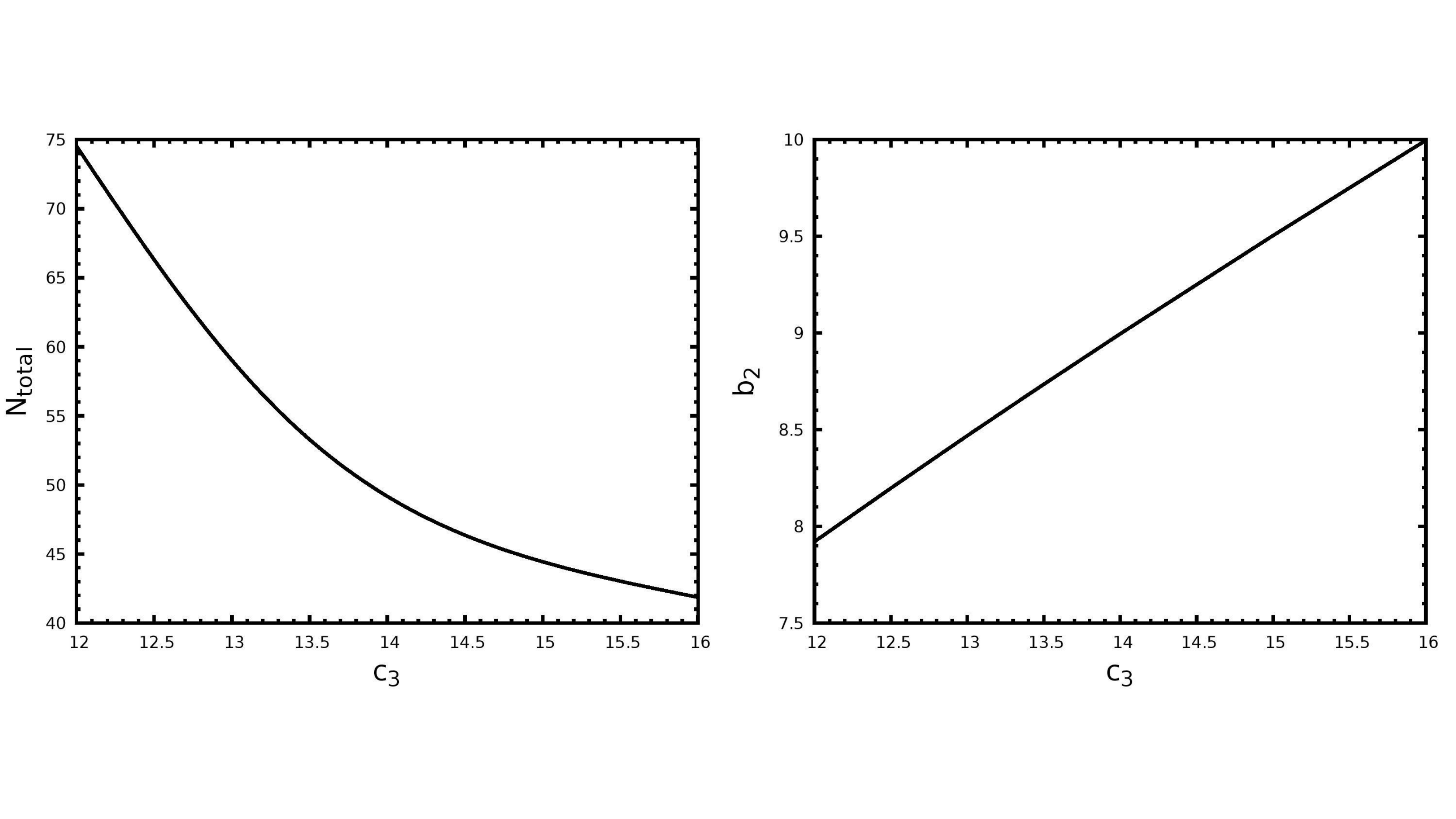}
\caption{Left: How parameters explain the total number of e-fold. Right: The dependence between the $c_3$ parameter and $b_2$.  }
\label{fdcb}
\end{figure}

 Furthermore,  the parameter $c_{1}$ and $c_3$ in  Tables~\ref{tkwz1} and  \ref{tkcc1} respectively are chosen in order to have the total number of e-fold between the two crossing condition in the range of $N_{total}=[45,55]$. For the case of modifying the K\"ahler potential, Eq.~(\ref{ciii1}), this feature is determined by the combination of parameters $\bar{\lambda}$ and $\bar{\mu}$. For instance, in Figure  \ref{fdcb} (left panel), it is illustrated the choice of parameter $c_3$  for the case (\ref{case2}). 
This parameter determines how sharp the power spectrum will be and then how small the range of e-fold in generation of PBH will be. This characteristic can be shown by comparing plots in Figure \ref{fbrapr1}, where the left panels have sharper peaks and their total number of e-folds is smaller than the right panels, and the corresponding plots in Figure~\ref{fbraabu1}. 
 As one can notice in Figure~\ref{fdcb} (right panel) the remaining parameter $b_i$ with $i=[1,3]$, which is found in all superpotentials and K\"ahler potential throughout this paper and is discussed below, has a strong dependence on the corresponding $c_{1,3,4}$. In this panel, we plot the dependence between the parameters $c_3$ and $b_2$ for case~ (\ref{case2}). By comparing the Table~\ref{tkcc1} and the Figure~\ref{fdcb} (right panel), it is easy to connect the number of e-fold with parameter $c_3$ and $b_2$. Hence, by demanding a number of e-fold we can deduce the parameter $c_3$ and then the parameter $b_2$.

\begin{figure}[h!]
\centering
\includegraphics[width=150mm]{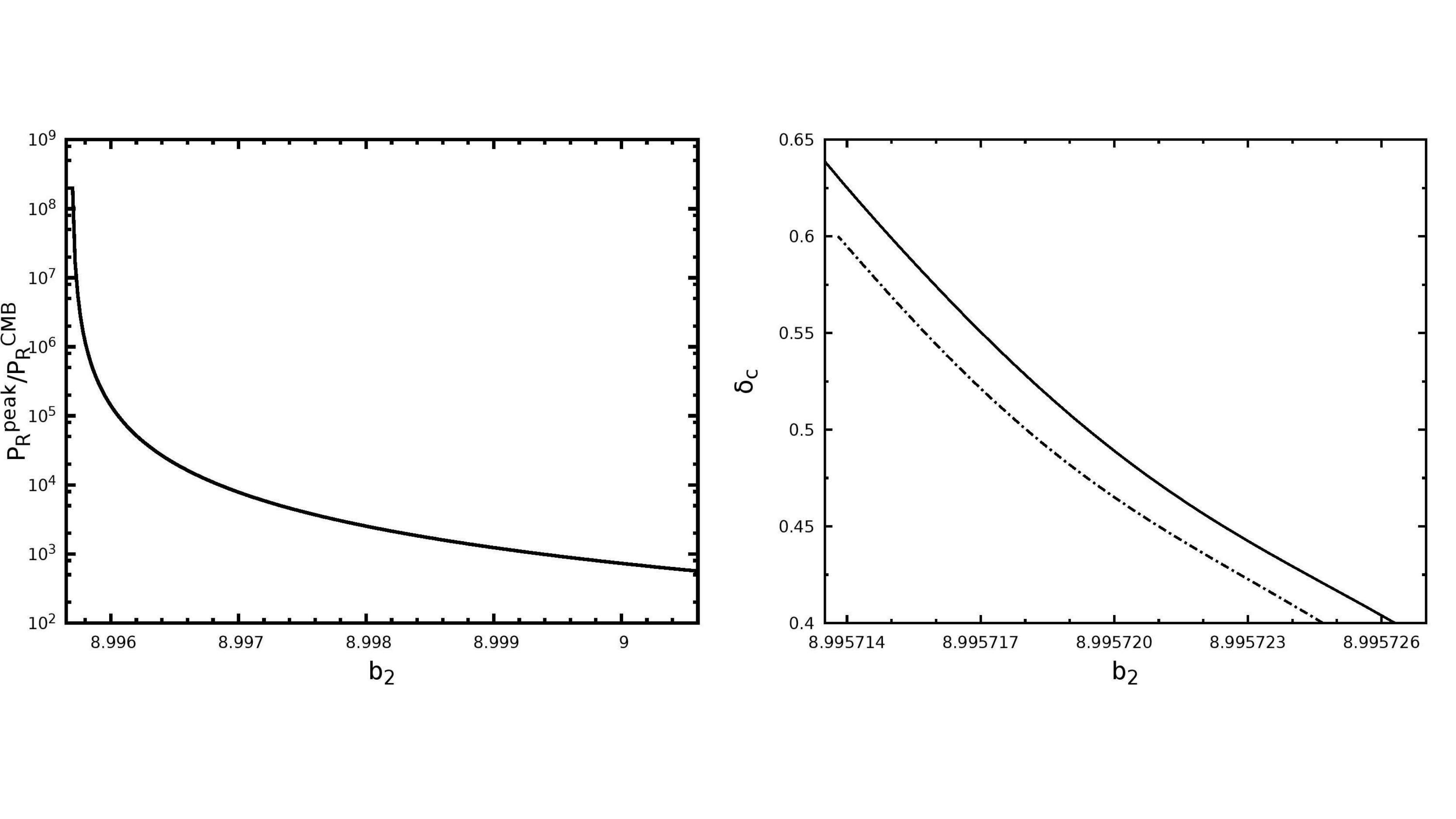}
\caption{How  parameter $b_i$  arises. Left: The dependence of $b_2$ to the enhancement of power spectrum.  Right: The dependence of $b_2$ to the $\delta_c$. Dashed lines corresponds to PS formalism and solid to PT for the same given value of $f_{PBH}=0.1$. }
\label{fb2}
\end{figure}

\par
 The parameter $b_i$, which demands more fine-tuning, is selected in order to achieve the proper  enhancement in power spectrum. In  Figure~\ref{fb2} (left) we depict the dependence of $b_2$ to this enhancement. In order to understand the fine-tuning issue of the parameter $b_i$, we evaluate the parameter $\Delta_{p}$, which is presented in Ref.~\cite{Barbieri:1987fn,Leggett:2014mza} and  is the maximum value of the following logarithmic derivative:
\begin{equation}
\Delta_{p}=\mathrm{Max} \left| \frac{\partial ln(f) }{\partial ln(x) }   \right|= \mathrm{Max} \left| \frac{\partial f }{\partial x } \frac{x}{f}  \right|
\label{deltaeenz}
\end{equation}
where we have assumed a general function $f(x)$.
Large values of $\Delta_{p}$ show that there is large fine-tuning. 
Taking this into account, we compute numerically the Eq.~ \ref{deltaeenz} for the function $P_R^{(peak)} (b)$ and we find that $\Delta_{p}  \sim 10^{6}$. Therefore, we notice that there is  a large fine-tuning in order to achieve significant peaks in power spectrum

\par 
Furthermore, the specific value of the parameter $b_i$ should be in accordance with the acceptable value of $\delta_c$ in order to have a significant abundance of PBH. This fact can be explained by the Figure \ref{fb2} (right), where we present the dependence of parameter $b_2$ by the $\delta_c$. It is has been remarked by previous work that the PBH abundances depend on the critical threshold density of collapse \cite{Ballesteros:2017fsr,Dalianis:2018frf}. This sensitivity is imprinted in Figure \ref{fb2}, as a lot of fine-tuning is required in order to retrieve abundances with an acceptable value of $\delta_c$. In our work we use constraints by Refs.~\cite{Musco:2018rwt,Escriva:2019phb} and by this demand we find the value of $b_i$. 
Concretely, in Figure \ref{fb2} we set fractional abundance $f_{PBH}$ to be equal to $0.1$ and by this we find what values of $b_2$ corresponds to given values of $\delta_c$. In this figure we plot the dependence of parameter $b_2$ from $\delta_c$ by using both PS approach (dashed line) and PT (solid lines), with the first set of parameters from Table~\ref{tkcc1}. One can observe that for a given $b_2$ and $f_{PBH}$, $\delta_c$ is higher in PT than in PS approach.

%
\section{Conclusions}
  \label{conclusion}

In this work we provide a class of scalar potentials in order to generate PBH which is derived by no-scale supergravity.  We study superpotentials using the non-compact SU(2,1)/SU(2)$\times$ U(1) symmetry from  a previous work \cite{Ellis:20181,Ellis:2013nxa}.  By a proper modification in these superpotential, we derive potentials which are capable to generate PBH and can obtain the same transformation laws as they do without this modification.  Specifically, in the case of ($y_1$,$y_2$) basis, the transformation laws are conserved. These potentials lead to Starobinsky-like scalar potential if we reduce the system by two chiral fields system to one single field and set the extra parameters to zero.
\par  
  As a second model we derive such potentials by modifying K\"ahler potential and keeping the superpotential unchanged.  In this model we break  SU(2,1)/SU(2)$\times$ U(1) symmetry by adding extra terms in the K\"ahler potential. Therefore, we derive a scalar potential with an inflection point, which remains flat for large values of inflaton. As before this potential yields to Starobinsky scalar potential when we discard the extra parameters.  In all these potentials, derived both by modifying K\"ahler potential and by modifying superpotential, we study the stabilization before proceed to the evaluation of the single field.
   
\par  
All class of scalar potentials, provided in this work,  lead to the observed spectral index $n_s$, the observed tensor-to-scalar ratio $r$, the observed amplitude of power spectrum on large scale, the proposed number of e-fold between the  end of inflation and the time that today's largest observable scales exit the Hubble radius and the proposed $\delta_c$ in order to find significant abundance. In all of our cases it is useful to prove that slow-roll approximation fails to produce the correct power spectrum in agreement with other works in literature \cite{Ballesteros:2017fsr,Cicoli:2018asa,Gao:2018pvq,Dalianis:2018frf,Mahbub:2019uhl}. Therefore, the numerical computation for perturbation is imperative.

Furthermore, in order to evaluate the fractional abundance of PBHs, we take into consideration two methods, the PS approach and the PT one, and we show their difference in our models. We find that there is a discrepancy between these approaches, which it can be eliminated by the uncertainly  in the critical value $\delta_c$.  

  \par
 As we have remarked the apparent drawback of such models is that a high level of  fine-tuning  is required, in order to achieve the desirable peaks in the  power spectrum. We deduce that we have the same issue of fine-tuning as it is analyzed in Ref.~\cite{Hertzberg:2017dkh}. 
Other theoretical approaches such as multifield inflation in hybrid models seems more natural \cite{Clesse:2015wea} and a complete study in PBH should mention to them. However, those models cannot predict the correct spectral index on CMB scales\cite{Halpern:2014mca}. Hence an alluring outcome study should provide a less fine-tuning model embedded in the framework of hybrid models.

\section*{Acknowledgments}
  \label{acknow}
The research work was supported by the Hellenic Foundation for Research
and Innovation (H.F.R.I.) under the “ First Call for H.F.R.I. Research Projects to support Faculty members and
Researchers and the procurement of high-cost research equipment grant ” (Project Number: 824). I.D.S. would like to thank V.C. Spanos for useful discussions.


\bibliographystyle{elsarticle-num}
\bibliography{bib}

\begin{thebibliography}{10}
\expandafter\ifx\csname url\endcsname\relax
  \def\url#1{\texttt{#1}}\fi
\expandafter\ifx\csname urlprefix\endcsname\relax\def\urlprefix{URL }\fi
\expandafter\ifx\csname href\endcsname\relax
  \def\href#1#2{#2} \def\path#1{#1}\fi

\bibitem{Abbott:2016blz}
B.~P. Abbott, et~al., {Observation of Gravitational Waves from a Binary Black
  Hole Merger}, Phys. Rev. Lett. 116~(6) (2016) 061102.
\newblock \href {http://arxiv.org/abs/1602.03837} {\path{arXiv:1602.03837}},
  \href {https://doi.org/10.1103/PhysRevLett.116.061102}
  {\path{doi:10.1103/PhysRevLett.116.061102}}.

\bibitem{Abbott:2016nmj}
B.~P. Abbott, et~al., {GW151226: Observation of Gravitational Waves from a
  22-Solar-Mass Binary Black Hole Coalescence}, Phys. Rev. Lett. 116~(24)
  (2016) 241103.
\newblock \href {http://arxiv.org/abs/1606.04855} {\path{arXiv:1606.04855}},
  \href {https://doi.org/10.1103/PhysRevLett.116.241103}
  {\path{doi:10.1103/PhysRevLett.116.241103}}.

\bibitem{Abbott:2017vtc}
B.~P. Abbott, et~al., {GW170104: Observation of a 50-Solar-Mass Binary Black
  Hole Coalescence at Redshift 0.2}, Phys. Rev. Lett. 118~(22) (2017) 221101,
  [Erratum: Phys.Rev.Lett. 121, 129901 (2018)].
\newblock \href {http://arxiv.org/abs/1706.01812} {\path{arXiv:1706.01812}},
  \href {https://doi.org/10.1103/PhysRevLett.118.221101}
  {\path{doi:10.1103/PhysRevLett.118.221101}}.

\bibitem{Abbott:2017gyy}
B.~P. Abbott, et~al., {GW170608: Observation of a 19-solar-mass Binary Black
  Hole Coalescence}, Astrophys. J. Lett. 851 (2017) L35.
\newblock \href {http://arxiv.org/abs/1711.05578} {\path{arXiv:1711.05578}},
  \href {https://doi.org/10.3847/2041-8213/aa9f0c}
  {\path{doi:10.3847/2041-8213/aa9f0c}}.

\bibitem{Abbott:2017oio}
B.~P. Abbott, et~al., {GW170814: A Three-Detector Observation of Gravitational
  Waves from a Binary Black Hole Coalescence}, Phys. Rev. Lett. 119~(14) (2017)
  141101.
\newblock \href {http://arxiv.org/abs/1709.09660} {\path{arXiv:1709.09660}},
  \href {https://doi.org/10.1103/PhysRevLett.119.141101}
  {\path{doi:10.1103/PhysRevLett.119.141101}}.

\bibitem{Ballesteros:2017fsr}
G.~Ballesteros, M.~Taoso, Primordial black hole dark matter from single field
  inflation, Physical Review D 97~(2) (jan 2018).
\newblock \href {https://doi.org/10.1103/physrevd.97.023501}
  {\path{doi:10.1103/physrevd.97.023501}}.

\bibitem{Ozsoy:2018flq}
O.~Özsoy, S.~Parameswaran, G.~Tasinato, I.~Zavala, Mechanisms for primordial
  black hole production in string theory, Journal of Cosmology and
  Astroparticle Physics 2018~(07) (2018) 005--005.
\newblock \href {https://doi.org/10.1088/1475-7516/2018/07/005}
  {\path{doi:10.1088/1475-7516/2018/07/005}}.

\bibitem{Biagetti:2018pjj}
M.~Biagetti, G.~Franciolini, A.~Kehagias, A.~Riotto, Primordial black holes
  from inflation and quantum diffusion, Journal of Cosmology and Astroparticle
  Physics 2018~(07) (2018) 032--032.
\newblock \href {https://doi.org/10.1088/1475-7516/2018/07/032}
  {\path{doi:10.1088/1475-7516/2018/07/032}}.

\bibitem{Franciolini:2018vbk}
G.~Franciolini, A.~Kehagias, S.~Matarrese, A.~Riotto, Primordial black holes
  from inflation and non-gaussianity, Journal of Cosmology and Astroparticle
  Physics 2018~(03) (2018) 016--016.
\newblock \href {https://doi.org/10.1088/1475-7516/2018/03/016}
  {\path{doi:10.1088/1475-7516/2018/03/016}}.

\bibitem{Gao:2018pvq}
T.-J. Gao, Z.-K. Guo, Primordial black hole production in inflationary models
  of supergravity with a single chiral superfield, Physical Review D 98~(6)
  (sep 2018).
\newblock \href {https://doi.org/10.1103/physrevd.98.063526}
  {\path{doi:10.1103/physrevd.98.063526}}.

\bibitem{Cicoli:2018asa}
M.~Cicoli, V.~A. Diaz, F.~G. Pedro, Primordial black holes from string
  inflation, Journal of Cosmology and Astroparticle Physics 2018~(06) (2018)
  034--034.
\newblock \href {https://doi.org/10.1088/1475-7516/2018/06/034}
  {\path{doi:10.1088/1475-7516/2018/06/034}}.

\bibitem{Dalianis:2018frf}
I.~Dalianis, A.~Kehagias, G.~Tringas, Primordial black holes from
  $\alpha$-attractors, Journal of Cosmology and Astroparticle Physics 2019~(01)
  (2019) 037--037.
\newblock \href {https://doi.org/10.1088/1475-7516/2019/01/037}
  {\path{doi:10.1088/1475-7516/2019/01/037}}.

\bibitem{Garcia-Bellido:2017mdw}
J.~Garc{\'{\i}}a-Bellido, E.~R. Morales, Primordial black holes from single
  field models of inflation, Physics of the Dark Universe 18 (2017) 47--54.
\newblock \href {https://doi.org/10.1016/j.dark.2017.09.007}
  {\path{doi:10.1016/j.dark.2017.09.007}}.

\bibitem{Ezquiaga:2017fvi}
J.~M. Ezquiaga, J.~Garc{\'{\i}}a-Bellido, E.~R. Morales, Primordial black hole
  production in critical higgs inflation, Physics Letters B 776 (2018)
  345--349.
\newblock \href {https://doi.org/10.1016/j.physletb.2017.11.039}
  {\path{doi:10.1016/j.physletb.2017.11.039}}.

\bibitem{Gong:2017qlj}
H.~Di, Y.~Gong, Primordial black holes and second order gravitational waves
  from ultra-slow-roll inflation, Journal of Cosmology and Astroparticle
  Physics 2018~(07) (2018) 007--007.
\newblock \href {https://doi.org/10.1088/1475-7516/2018/07/007}
  {\path{doi:10.1088/1475-7516/2018/07/007}}.

\bibitem{Hertzberg:2017dkh}
M.~P. Hertzberg, M.~Yamada, Primordial black holes from polynomial potentials
  in single field inflation, Physical Review D 97~(8) (apr 2018).
\newblock \href {https://doi.org/10.1103/physrevd.97.083509}
  {\path{doi:10.1103/physrevd.97.083509}}.

\bibitem{Espinosa:2017sgp}
J.~Espinosa, D.~Racco, A.~Riotto, Cosmological signature of the standard model
  higgs vacuum instability: Primordial black holes as dark matter, Physical
  Review Letters 120~(12) (mar 2018).
\newblock \href {https://doi.org/10.1103/physrevlett.120.121301}
  {\path{doi:10.1103/physrevlett.120.121301}}.

\bibitem{Clesse:2015wea}
S.~Clesse, J.~Garc{\'{\i}}a-Bellido, Massive primordial black holes from hybrid
  inflation as dark matter and the seeds of galaxies, Physical Review D 92~(2)
  (jul 2015).
\newblock \href {https://doi.org/10.1103/physrevd.92.023524}
  {\path{doi:10.1103/physrevd.92.023524}}.

\bibitem{Germani:2017bcs}
C.~Germani, T.~Prokopec, On primordial black holes from an inflection point,
  Phys. Dark Universe 18 (2017) 6--10.
\newblock \href {https://doi.org/10.1016/j.dark.2017.09.001}
  {\path{doi:10.1016/j.dark.2017.09.001}}.

\bibitem{Inomata:2017okj}
K.~Inomata, M.~Kawasaki, K.~Mukaida, Y.~Tada, T.~T. Yanagida, Inflationary
  primordial black holes as all dark matter, Physical Review D 96~(4) (aug
  2017).
\newblock \href {https://doi.org/10.1103/physrevd.96.043504}
  {\path{doi:10.1103/physrevd.96.043504}}.

\bibitem{Motohashi:2017kbs}
H.~Motohashi, W.~Hu, Primordial black holes and slow-roll violation, Physical
  Review D 96~(6) (sep 2017).
\newblock \href {https://doi.org/10.1103/physrevd.96.063503}
  {\path{doi:10.1103/physrevd.96.063503}}.

\bibitem{Kannike:2017bxn}
K.~Kannike, L.~Marzola, M.~Raidal, H.~Veermäe, Single field double inflation
  and primordial black holes, J. Cosmol. Astropart. Phys. 2017~(09) (2017)
  020--020.
\newblock \href {https://doi.org/10.1088/1475-7516/2017/09/020}
  {\path{doi:10.1088/1475-7516/2017/09/020}}.

\bibitem{Inomata:2017vxo}
K.~Inomata, M.~Kawasaki, K.~Mukaida, T.~T. Yanagida, Double inflation as a
  single origin of primordial black holes for all dark matter and {LIGO}
  observations, Physical Review D 97~(4) (feb 2018).
\newblock \href {https://doi.org/10.1103/physrevd.97.043514}
  {\path{doi:10.1103/physrevd.97.043514}}.

\bibitem{Kawasaki:2016pql}
M.~Kawasaki, A.~Kusenko, Y.~Tada, T.~T. Yanagida, Primordial black holes as
  dark matter in supergravity inflation models, Physical Review D 94~(8) (oct
  2016).
\newblock \href {https://doi.org/10.1103/physrevd.94.083523}
  {\path{doi:10.1103/physrevd.94.083523}}.

\bibitem{Carr:2016drx}
B.~Carr, F.~Kühnel, M.~Sandstad, Primordial black holes as dark matter,
  Physical Review D 94~(8) (oct 2016).
\newblock \href {https://doi.org/10.1103/physrevd.94.083504}
  {\path{doi:10.1103/physrevd.94.083504}}.

\bibitem{Cheng:2018yyr}
S.-L. Cheng, W.~Lee, K.-W. Ng, Primordial black holes and associated
  gravitational waves in axion monodromy inflation, J. Cosmol. Astropart. Phys.
  2018~(07) (2018) 001--001.
\newblock \href {https://doi.org/10.1088/1475-7516/2018/07/001}
  {\path{doi:10.1088/1475-7516/2018/07/001}}.

\bibitem{Mahbub:2019uhl}
R.~Mahbub, {Primordial black hole formation in inflationary $\alpha$-attractor
  models}, Phys. Rev. D101~(2) (2020) 023533.
\newblock \href {http://arxiv.org/abs/1910.10602} {\path{arXiv:1910.10602}},
  \href {https://doi.org/10.1103/PhysRevD.101.023533}
  {\path{doi:10.1103/PhysRevD.101.023533}}.

\bibitem{Mishra19}
S.~S. Mishra, V.~Sahni, {Primordial Black Holes from a tiny bump/dip in the
  Inflaton potential}, JCAP 04 (2020) 007.
\newblock \href {http://arxiv.org/abs/1911.00057} {\path{arXiv:1911.00057}},
  \href {https://doi.org/10.1088/1475-7516/2020/04/007}
  {\path{doi:10.1088/1475-7516/2020/04/007}}.

\bibitem{Ballesteros:2020qam}
G.~Ballesteros, J.~Rey, M.~Taoso, A.~Urbano, {Primordial black holes as dark
  matter and gravitational waves from single-field polynomial inflation}
  (2020).
\newblock \href {http://arxiv.org/abs/2001.08220} {\path{arXiv:2001.08220}}.

\bibitem{Ballesteros:2019hus}
G.~Ballesteros, J.~Rey, F.~Rompineve, {Detuning primordial black hole dark
  matter with early matter domination and axion monodromy} (2019).
\newblock \href {http://arxiv.org/abs/1912.01638} {\path{arXiv:1912.01638}}.

\bibitem{Nanopoulos:2020nnh}
D.~V. Nanopoulos, V.~C. Spanos, I.~D. Stamou, {Primordial Black Holes from
  No-Scale Supergravity}, Phys. Rev. D 102~(8) (2020) 083536.
\newblock \href {http://arxiv.org/abs/2008.01457} {\path{arXiv:2008.01457}},
  \href {https://doi.org/10.1103/PhysRevD.102.083536}
  {\path{doi:10.1103/PhysRevD.102.083536}}.

\bibitem{Braglia:2020eai}
M.~Braglia, D.~K. Hazra, F.~Finelli, G.~F. Smoot, L.~Sriramkumar, A.~A.
  Starobinsky, {Generating PBHs and small-scale GWs in two-field models of
  inflation}, JCAP 08 (2020) 001.
\newblock \href {http://arxiv.org/abs/2005.02895} {\path{arXiv:2005.02895}},
  \href {https://doi.org/10.1088/1475-7516/2020/08/001}
  {\path{doi:10.1088/1475-7516/2020/08/001}}.

\bibitem{Aldabergenov:2020bpt}
Y.~Aldabergenov, A.~Addazi, S.~V. Ketov, {Primordial black holes from modified
  supergravity} (7 2020).
\newblock \href {http://arxiv.org/abs/2006.16641} {\path{arXiv:2006.16641}}.

\bibitem{Anguelova:2020nzl}
L.~Anguelova, {On Primordial Black Holes from Rapid Turns in Two-field Models}
  (12 2020).
\newblock \href {http://arxiv.org/abs/2012.03705} {\path{arXiv:2012.03705}}.

\bibitem{Braglia:2020taf}
M.~Braglia, X.~Chen, D.~K. Hazra, {Probing Primordial Features with the
  Stochastic Gravitational Wave Background} (12 2020).
\newblock \href {http://arxiv.org/abs/2012.05821} {\path{arXiv:2012.05821}}.

\bibitem{Ade:2015lrj}
B.~A. et~al. (LIGO Scientific~Collaboration, V.~Collaboration), Planck2015
  results, Astronomy {\&} Astrophysics 594 (2016) A20.
\newblock \href {https://doi.org/10.1051/0004-6361/201525898}
  {\path{doi:10.1051/0004-6361/201525898}}.

\bibitem{Akrami:2018odb}
Y.~A. et.~al. [Planck~Collaboration], Planck 2018 results. x. constraints on
  inflation\href {http://arxiv.org/abs/http://arxiv.org/abs/1807.06211v2}
  {\path{arXiv:http://arxiv.org/abs/1807.06211v2}}.

\bibitem{Ellis:1982a}
E.~J.R., D.~V. Nanopoulos, O.~K.A, T.~K., {Cosmological Inflation Cries Out for
  Supersymmetry}, Phys.Lett.B 118 (1982) 335--339.
\newblock \href {https://doi.org/10.1016/0370-2693(82)90198-8}
  {\path{doi:10.1016/0370-2693(82)90198-8}}.

\bibitem{Ellis:2013nxa}
J.~Ellis, D.~V. Nanopoulos, K.~A. Olive, {Starobinsky-like Inflationary Models
  as Avatars of No-Scale Supergravity}, JCAP 1310 (2013) 009.
\newblock \href {http://arxiv.org/abs/1307.3537} {\path{arXiv:1307.3537}},
  \href {https://doi.org/10.1088/1475-7516/2013/10/009}
  {\path{doi:10.1088/1475-7516/2013/10/009}}.

\bibitem{Cremer:1983}
E.~Cremmer, S.~Ferrara, C.~Kounas, D.~V. Nanopoulos, {Naturally vanishing
  cosmological constant in N=1 supergravity}, Phys.Lett.B 133 (1983) 61--66.
\newblock \href {https://doi.org/10.1016/0370-2693(83)90106-5}
  {\path{doi:10.1016/0370-2693(83)90106-5}}.

\bibitem{Ellislahanas:1984}
J.~Ellis, A.~Lahanas, D.~V. Nanopoulos, K.~Tamvakis, {No-Scale Supersymmetric
  Standard Model}, Phys.Lett.B 134 (1984) 429--435.
\newblock \href {https://doi.org/10.1016/0370-2693(84)91378-9}
  {\path{doi:10.1016/0370-2693(84)91378-9}}.

\bibitem{Elliskounas:1984a}
J.~R. Ellis, C.~Kounnas, D.~V. Nanopoulos, {Phenomenological SU(1,1)
  Supergravity}, Nucl. Phys. B 241 (1984) 406--428.
\newblock \href {https://doi.org/10.1016/0550-3213(84)90054-3}
  {\path{doi:10.1016/0550-3213(84)90054-3}}.

\bibitem{Elliskounas:1984b}
J.~R. Ellis, C.~Kounnas, D.~V. Nanopoulos, {No Scale Supersymmetric Guts},
  Nucl. Phys. B 247 (1984) 373--395.
\newblock \href {https://doi.org/10.1016/0550-3213(84)90555-8}
  {\path{doi:10.1016/0550-3213(84)90555-8}}.

\bibitem{Lahanas:1986}
A.~B. Lahanas, D.~V. Nanopoulos, {The Road to No Scale Supergravity}, Phys.
  Rept. 145 (1987) 1.
\newblock \href {https://doi.org/10.1016/0370-1573(87)90034-2}
  {\path{doi:10.1016/0370-1573(87)90034-2}}.

\bibitem{Ellis:20181}
J.~Ellis, D.~V. Nanopoulos, K.~A. Olive, S.~Verner, {A general classification
  of Starobinsky-like inflationary avatars of SU(2,1)/SU(2) $\times$ U(1)
  no-scale supergravity}, JHEP 03 (2019) 099.
\newblock \href {http://arxiv.org/abs/1812.02192} {\path{arXiv:1812.02192}},
  \href {https://doi.org/10.1007/JHEP03(2019)099}
  {\path{doi:10.1007/JHEP03(2019)099}}.

\bibitem{Press74}
W.~H. Press, P.~Schechter, Formation of galaxies and clusters of galaxies by
  self-similar gravitational condensation, The Astrophysical Journal 187 (1974)
  425.
\newblock \href {https://doi.org/10.1086/152650} {\path{doi:10.1086/152650}}.

\bibitem{Bardeen:1985tr}
J.~M. Bardeen, J.~Bond, N.~Kaiser, A.~Szalay, {The Statistics of Peaks of
  Gaussian Random Fields}, Astrophys. J. 304 (1986) 15--61.
\newblock \href {https://doi.org/10.1086/164143} {\path{doi:10.1086/164143}}.

\bibitem{Ellis:2013xoa}
J.~Ellis, D.~V. Nanopoulos, K.~A. Olive, Publisher's note: No-scale
  supergravity realization of the starobinsky model of inflation [phys. rev.
  lett.111, 111301 (2013)], Physical Review Letters 111~(12) (sep 2013).
\newblock \href {https://doi.org/10.1103/physrevlett.111.129902}
  {\path{doi:10.1103/physrevlett.111.129902}}.

\bibitem{Cecotti:1987a}
C.~S., {Higher derivative supergravity is equivalent to standard supergravity
  coupled to matter}, Phys.Lett.B 190 (1987) 86--92.
\newblock \href {https://doi.org/10.1016/0370-2693(87)90844-6}
  {\path{doi:10.1016/0370-2693(87)90844-6}}.

\bibitem{Ellis:2019bmm}
J.~Ellis, D.~V. Nanopoulos, K.~A. Olive, S.~Verner, Unified no-scale
  attractors, Journal of Cosmology and Astroparticle Physics 2019~(09) (2019)
  040--040.
\newblock \href {https://doi.org/10.1088/1475-7516/2019/09/040}
  {\path{doi:10.1088/1475-7516/2019/09/040}}.

\bibitem{Kallosh:2010ug}
R.~Kallosh, A.~Linde, New models of chaotic inflation in supergravity, Journal
  of Cosmology and Astroparticle Physics 2010~(11) (2010) 011--011.
\newblock \href {https://doi.org/10.1088/1475-7516/2010/11/011}
  {\path{doi:10.1088/1475-7516/2010/11/011}}.

\bibitem{Kallosh:2013xya}
R.~Kallosh, A.~Linde, {Superconformal generalizations of the Starobinsky
  model}, JCAP 1306 (2013) 028.
\newblock \href {http://arxiv.org/abs/1306.3214} {\path{arXiv:1306.3214}},
  \href {https://doi.org/10.1088/1475-7516/2013/06/028}
  {\path{doi:10.1088/1475-7516/2013/06/028}}.

\bibitem{WessBagger}
J.~Wess, J.~Bagger, {Superconformal generalizations of the Starobinsky model}
  (1992).

\bibitem{Akrami2018}
Y.~A. et~al. [Planck~Collaboration], Planck 2018 results. vi. cosmological
  parameters\href {http://arxiv.org/abs/http://arxiv.org/abs/1807.06209v2}
  {\path{arXiv:http://arxiv.org/abs/1807.06209v2}}.

\bibitem{Stewart:1993bc}
E.~D. Stewart, D.~H. Lyth, {A More accurate analytic calculation of the
  spectrum of cosmological perturbations produced during inflation}, Phys.
  Lett. B302 (1993) 171--175.
\newblock \href {http://arxiv.org/abs/gr-qc/9302019}
  {\path{arXiv:gr-qc/9302019}}, \href
  {https://doi.org/10.1016/0370-2693(93)90379-V}
  {\path{doi:10.1016/0370-2693(93)90379-V}}.

\bibitem{Ringeval:2007am}
C.~Ringeval, {The exact numerical treatment of inflationary models}, Lect.
  Notes Phys. 738 (2008) 243--273.
\newblock \href {http://arxiv.org/abs/astro-ph/0703486}
  {\path{arXiv:astro-ph/0703486}}, \href
  {https://doi.org/10.1007/978-3-540-74353-8_7}
  {\path{doi:10.1007/978-3-540-74353-8_7}}.

\bibitem{Bunch:1978yq}
T.~Bunch, P.~Davies, {Quantum Field Theory in de Sitter Space: Renormalization
  by Point Splitting}, Proc. Roy. Soc. Lond. A 360 (1978) 117--134.
\newblock \href {https://doi.org/10.1098/rspa.1978.0060}
  {\path{doi:10.1098/rspa.1978.0060}}.

\bibitem{Young:2014ana}
S.~Young, C.~T. Byrnes, M.~Sasaki, {Calculating the mass fraction of primordial
  black holes}, JCAP 07 (2014) 045.
\newblock \href {http://arxiv.org/abs/1405.7023} {\path{arXiv:1405.7023}},
  \href {https://doi.org/10.1088/1475-7516/2014/07/045}
  {\path{doi:10.1088/1475-7516/2014/07/045}}.

\bibitem{Green:2004wb}
A.~M. Green, A.~R. Liddle, K.~A. Malik, M.~Sasaki, {A New calculation of the
  mass fraction of primordial black holes}, Phys. Rev. D 70 (2004) 041502.
\newblock \href {http://arxiv.org/abs/astro-ph/0403181}
  {\path{arXiv:astro-ph/0403181}}, \href
  {https://doi.org/10.1103/PhysRevD.70.041502}
  {\path{doi:10.1103/PhysRevD.70.041502}}.

\bibitem{Carr1975}
B.~J. Carr, The primordial black hole mass spectrum, The Astrophysical Journal
  201 (1975) 1.
\newblock \href {https://doi.org/10.1086/153853} {\path{doi:10.1086/153853}}.

\bibitem{Harada:2013epa}
T.~Harada, C.-M. Yoo, K.~Kohri, Threshold of primordial black hole formation,
  Physical Review D 88~(8) (oct 2013).
\newblock \href {https://doi.org/10.1103/physrevd.88.084051}
  {\path{doi:10.1103/physrevd.88.084051}}.

\bibitem{Musco:2008hv}
I.~Musco, J.~C. Miller, A.~G. Polnarev, Primordial black hole formation in the
  radiative era: investigation of the critical nature of the collapse,
  Classical and Quantum Gravity 26~(23) (2009) 235001.
\newblock \href {https://doi.org/10.1088/0264-9381/26/23/235001}
  {\path{doi:10.1088/0264-9381/26/23/235001}}.

\bibitem{Musco:2004ak}
I.~Musco, J.~C. Miller, L.~Rezzolla, Computations of primordial black-hole
  formation, Classical and Quantum Gravity 22~(7) (2005) 1405--1424.
\newblock \href {https://doi.org/10.1088/0264-9381/22/7/013}
  {\path{doi:10.1088/0264-9381/22/7/013}}.

\bibitem{Musco:2012au}
I.~Musco, J.~C. Miller, Primordial black hole formation in the early universe:
  critical behaviour and self-similarity, Classical and Quantum Gravity 30~(14)
  (2013) 145009.
\newblock \href {https://doi.org/10.1088/0264-9381/30/14/145009}
  {\path{doi:10.1088/0264-9381/30/14/145009}}.

\bibitem{Musco:2018rwt}
I.~Musco, {Threshold for primordial black holes: Dependence on the shape of the
  cosmological perturbations}, Phys. Rev. D 100~(12) (2019) 123524.
\newblock \href {http://arxiv.org/abs/1809.02127} {\path{arXiv:1809.02127}},
  \href {https://doi.org/10.1103/PhysRevD.100.123524}
  {\path{doi:10.1103/PhysRevD.100.123524}}.

\bibitem{Escriva:2019phb}
A.~Escrivà, C.~Germani, R.~K. Sheth, {Universal threshold for primordial black
  hole formation}, Phys. Rev. D 101~(4) (2020) 044022.
\newblock \href {http://arxiv.org/abs/1907.13311} {\path{arXiv:1907.13311}},
  \href {https://doi.org/10.1103/PhysRevD.101.044022}
  {\path{doi:10.1103/PhysRevD.101.044022}}.

\bibitem{Escriva:2020tak}
A.~Escrivà, C.~Germani, R.~K. Sheth, {Analytical thresholds for black hole
  formation in general cosmological backgrounds} (7 2020).
\newblock \href {http://arxiv.org/abs/2007.05564} {\path{arXiv:2007.05564}}.

\bibitem{Carr:2009jm}
B.~J. Carr, K.~Kohri, Y.~Sendouda, J.~Yokoyama, New cosmological constraints on
  primordial black holes, Physical Review D 81~(10) (may 2010).
\newblock \href {https://doi.org/10.1103/physrevd.81.104019}
  {\path{doi:10.1103/physrevd.81.104019}}.

\bibitem{Inoue:2017csr}
Y.~Inoue, A.~Kusenko, New x-ray bound on density of primordial black holes,
  Journal of Cosmology and Astroparticle Physics 2017~(10) (2017) 034--034.
\newblock \href {https://doi.org/10.1088/1475-7516/2017/10/034}
  {\path{doi:10.1088/1475-7516/2017/10/034}}.

\bibitem{Montero-Camacho:2019jte}
P.~Montero-Camacho, X.~Fang, G.~Vasquez, M.~Silva, C.~M. Hirata, {Revisiting
  constraints on asteroid-mass primordial black holes as dark matter
  candidates}, JCAP 08 (2019) 031.
\newblock \href {http://arxiv.org/abs/1906.05950} {\path{arXiv:1906.05950}},
  \href {https://doi.org/10.1088/1475-7516/2019/08/031}
  {\path{doi:10.1088/1475-7516/2019/08/031}}.

\bibitem{Katz:2018zrn}
A.~Katz, J.~Kopp, S.~Sibiryakov, W.~Xue, {Femtolensing by Dark Matter
  Revisited}, JCAP 12 (2018) 005.
\newblock \href {http://arxiv.org/abs/1807.11495} {\path{arXiv:1807.11495}},
  \href {https://doi.org/10.1088/1475-7516/2018/12/005}
  {\path{doi:10.1088/1475-7516/2018/12/005}}.

\bibitem{Poulin:2017bwe}
V.~Poulin, P.~D. Serpico, F.~Calore, S.~Clesse, K.~Kohri, {CMB} bounds on
  disk-accreting massive primordial black holes, Physical Review D 96~(8) (oct
  2017).
\newblock \href {https://doi.org/10.1103/physrevd.96.083524}
  {\path{doi:10.1103/physrevd.96.083524}}.

\bibitem{Capela:2013yf}
F.~Capela, M.~Pshirkov, P.~Tinyakov, Constraints on primordial black holes as
  dark matter candidates from capture by neutron stars, Physical Review D
  87~(12) (jun 2013).
\newblock \href {https://doi.org/10.1103/physrevd.87.123524}
  {\path{doi:10.1103/physrevd.87.123524}}.

\bibitem{Niikura:2017zjd}
H.~Niikura, M.~Takada, N.~Yasuda, R.~H. Lupton, T.~Sumi, S.~More, T.~Kurita,
  S.~Sugiyama, A.~More, M.~Oguri, M.~Chiba, Microlensing constraints on
  primordial black holes with subaru/{HSC} andromeda observations, Nature
  Astronomy 3~(6) (2019) 524--534.
\newblock \href {https://doi.org/10.1038/s41550-019-0723-1}
  {\path{doi:10.1038/s41550-019-0723-1}}.

\bibitem{Wyrzykowski:2011tr}
L.~Wyrzykowski, et~al., {The OGLE View of Microlensing towards the Magellanic
  Clouds. IV. OGLE-III SMC Data and Final Conclusions on MACHOs}, Mon. Not.
  Roy. Astron. Soc. 416 (2011) 2949.
\newblock \href {http://arxiv.org/abs/1106.2925} {\path{arXiv:1106.2925}},
  \href {https://doi.org/10.1111/j.1365-2966.2011.19243.x}
  {\path{doi:10.1111/j.1365-2966.2011.19243.x}}.

\bibitem{Griest:2013esa}
K.~Griest, A.~M. Cieplak, M.~J. Lehner, {New Limits on Primordial Black Hole
  Dark Matter from an Analysis of Kepler Source Microlensing Data}, Phys. Rev.
  Lett. 111~(18) (2013) 181302.
\newblock \href {https://doi.org/10.1103/PhysRevLett.111.181302}
  {\path{doi:10.1103/PhysRevLett.111.181302}}.

\bibitem{Tisserand:2006zx}
P.~T. et. al., Limits on the macho content of the galactic halo from the
  {EROS}-2 survey of the magellanic clouds, Astronomy {\&} Astrophysics 469~(2)
  (2007) 387--404.
\newblock \href {https://doi.org/10.1051/0004-6361:20066017}
  {\path{doi:10.1051/0004-6361:20066017}}.

\bibitem{Ali-Haimoud:2016mbv}
Y.~Ali-Haïmoud, M.~Kamionkowski, Cosmic microwave background limits on
  accreting primordial black holes, Physical Review D 95~(4) (feb 2017).
\newblock \href {https://doi.org/10.1103/physrevd.95.043534}
  {\path{doi:10.1103/physrevd.95.043534}}.

\bibitem{Gaggero:2016dpq}
D.~Gaggero, G.~Bertone, F.~Calore, R.~M. Connors, M.~Lovell, S.~Markoff,
  E.~Storm, Searching for primordial black holes in the radio and x-ray sky,
  Physical Review Letters 118~(24) (jun 2017).
\newblock \href {https://doi.org/10.1103/physrevlett.118.241101}
  {\path{doi:10.1103/physrevlett.118.241101}}.

\bibitem{Kallosh:2010xz}
R.~Kallosh, A.~Linde, T.~Rube, General inflaton potentials in supergravity,
  Physical Review D 83~(4) (feb 2011).
\newblock \href {https://doi.org/10.1103/physrevd.83.043507}
  {\path{doi:10.1103/physrevd.83.043507}}.

\bibitem{Ellis:2018xdr}
J.~Ellis, B.~Nagaraj, D.~V. Nanopoulos, K.~A. Olive, De sitter vacua in
  no-scale supergravity, Journal of High Energy Physics 2018~(11) (nov 2018).
\newblock \href {https://doi.org/10.1007/jhep11(2018)110}
  {\path{doi:10.1007/jhep11(2018)110}}.

\bibitem{Hawking1975}
S.~W. HAWKING, Black hole explosions?, Nature 248~(5443) (1974) 30--31.
\newblock \href {https://doi.org/10.1038/248030a0}
  {\path{doi:10.1038/248030a0}}.

\bibitem{Ricotti:2007au}
M.~Ricotti, J.~P. Ostriker, K.~J. Mack, {Effect of Primordial Black Holes on
  the Cosmic Microwave Background and Cosmological Parameter Estimates},
  Astrophys. J. 680 (2008) 829.
\newblock \href {http://arxiv.org/abs/0709.0524} {\path{arXiv:0709.0524}},
  \href {https://doi.org/10.1086/587831} {\path{doi:10.1086/587831}}.

\bibitem{Laha:2020ivk}
R.~Laha, J.~B. Muñoz, T.~R. Slatyer, {INTEGRAL constraints on primordial black
  holes and particle dark matter}, Phys. Rev. D 101~(12)  123514.
\newblock \href {http://arxiv.org/abs/2004.00627} {\path{arXiv:2004.00627}},
  \href {https://doi.org/10.1103/PhysRevD.101.123514}
  {\path{doi:10.1103/PhysRevD.101.123514}}.

\bibitem{Barbieri:1987fn}
R.~Barbieri, G.~Giudice, {Upper Bounds on Supersymmetric Particle Masses},
  Nucl. Phys. B 306 (1988) 63--76.
\newblock \href {https://doi.org/10.1016/0550-3213(88)90171-X}
  {\path{doi:10.1016/0550-3213(88)90171-X}}.

\bibitem{Leggett:2014mza}
T.~Leggett, T.~Li, J.~A. Maxin, D.~V. Nanopoulos, J.~W. Walker, {No Naturalness
  or Fine-tuning Problems from No-Scale Supergravity} (3 2014).
\newblock \href {http://arxiv.org/abs/1403.3099} {\path{arXiv:1403.3099}}.

\bibitem{Halpern:2014mca}
I.~F. Halpern, M.~P. Hertzberg, M.~A. Joss, E.~I. Sfakianakis, {A Density Spike
  on Astrophysical Scales from an N-Field Waterfall Transition}, Phys. Lett. B
  748 (2015) 132--143.
\newblock \href {http://arxiv.org/abs/1410.1878} {\path{arXiv:1410.1878}},
  \href {https://doi.org/10.1016/j.physletb.2015.06.076}
  {\path{doi:10.1016/j.physletb.2015.06.076}}.

\end{thebibliography}

\end{document}